\documentclass{mn2e}
\usepackage{epsfig}
\usepackage{txfonts}

\graphicspath{{figures/}}

\begin{document}

\title[V404 Cyg]
{
Comprehensive coverage of particle acceleration and kinetic feedback from the stellar mass black hole V404 Cygni}
\author[Fender]
       {R. P. Fender$^{1,2}$\thanks{email: rob.fender@physics.ox.ac.uk}, K. P. Mooley$^{1,3,4}$, S. E. Motta$^{1,5}$, J. S. Bright$^{1,6}$, D. R. A. Williams$^{1,7}$       
       \newauthor A. P. Rushton$^1$, R. J. Beswick$^7$,     J. C. A. Miller-Jones$^8$,  M. Kimura$^9$,  K.Isogai$^{10,11}$, \newauthor T. Kato$^{12}$           
       \\
       $^1$Astrophysics, Department of Physics, University of Oxford, Keble Road, Oxford OX1 3RH\\
       $^2$Department of Astronomy, University of Cape Town, Private Bag X3, Rondebosch 7701, South Africa\\
       $^3$National Radio Astronomy Observatory, Socorro, New Mexico 87801, USA \\
       $^4$Cahill Center for Astronomy and Astrophysics, MC 249-17 California Institute of Technology, Pasadena, CA 91125, USA\\
       $^5$Istituto Nazionale di Astrofisica, Osservatorio Astronomico di Brera, via E. Bianchi 46, 23807 Merate (LC), Italy\\
       $^6$Department of Physics and Astronomy, Northwestern University,
Evanston, IL 60208, USA\\
       $^7$MERLIN/VLBI National Facility, Jodrell Bank Observatory, The University of Manchester, Macclesfield, Cheshire, SK11 9DL\\
       $^8$International Centre for Radio Astronomy Research–Curtin University, Perth, WA 6845, Australia\\
       $^9$Cluster for Pioneering Research, Institute of Physical and Chemical Research (RIKEN), 2-1 Hirosawa, Wako, Saitama 351-0198, Japan\\
   $^{10}$Okayama Observatory, Kyoto University, 3037-5 Honjo, Kamogatacho, Asakuchi, Okayama 719-0232, Japan\\
   $^{11}$Department of Multi-Disciplinary Sciences, Graduate School of Arts and Sciences, The University of Tokyo, 3-8-1 Komaba, Meguro, Tokyo 153-8902, Japan\\
  $^{12}$Department of Astronomy, Graduate School of Science, Kyoto University, Oiwakecho, Kitashirakawa, Sakyo-ku, Kyoto, Kyoto 606-8502, Japan
      }
\maketitle
\begin{abstract}
We present analysis of comprehensive radio observations of the black hole V404 Cyg during its 2015 outburst. These data represent the best ever coverage of jet production and particle acceleration from any black hole. We report for the first time a clear and near-linear flux-rms correlation in the radio flux densities. Investigation of individual flares reveals in nearly all cases the peak corresponds to the transition from optically thick to thin to synchrotron emission, but an extended phase of particle acceleration is required in contrast to simple impulsive injection models. The largest radio flare is preceded by a phase of optical oscillations and followed one day later by a smaller but optically thin flare, likely due to ejecta interacting with the interstellar medium. Comparing the radio emission to contemporaneous X-ray and optical data, we find that the X-ray and radio measurements are correlated on all timescales from seconds to one day. Correlation with the optical flux densities is weak at short timescales, but becomes significant on timescales greater than a few hours. We evaluate the physical conditions (size, magnetic field and internal energy) associated with 86 individual radio flares, which in turn allows us to place a lower limit on the kinetic feedback over the 15 days of intense activity. If this energy was deposited locally to the source, as implied by the failure to detect jets on angular scales larger than milliarcsec, then we predict that a nova-like shell could have been formed. 

\end{abstract}

\section{Introduction}

Understanding how black hole accretion proceeds is a fundamental goal for observational astrophysics. A crucial aspect in this research is determing how and in what form a large fraction of the available gravitational potential energy is fed back to the surrounding environment via radiation, winds and jets (e.g. McNamara \& Nulsen 2007; Fender \& Mu\~noz-Darias 2016 and references therein). Studies of `stellar mass' (typically $M \sim 7 M_{\odot}$) black holes accreting from a companion star in X-ray binary systems have led to a clear phenomenological picture of how accretion rates and modes, together known as `states', connect to the power and mode of kinetic feedback in most black holes (Fender, Belloni \& Gallo 2004; Ponti et al. 2012). This picture currently seems to be broadly consistent with what is seen on much larger scales in the accreting supermassive black holes at the centres of galaxies, the active galactic nuclei (K\"ording, Jester \& Fender 2006; Svoboda, Guainazzi \& Merloni 2017; Fernandez-Ontiveros \& Munoz-Darias 2021). Feedback from AGN seems likely to have had an impact on the growth of the largest galaxies (McNamara \& Nulsen 2007), and thus studies of black hole X-ray binaries inform us not only about the physics of accretion but about the growth of galaxies over cosmological time. A wide range of other explosive and extreme phenomena are also associated with jets from black holes (e.g. Tidal Disruption Events [Zauderer et al. 2009], Gamma Ray Bursts [Gehrels, Ramirez-Ruiz \& Fox 2009], merging neutron stars [Hallinan et al. 2017]) and it may be hoped that our studies can shed light on all of these phenomena. The empirical framework connecting accretion states, jets and winds in black hole X-ray binaries, summarised in Fender \& Belloni (2012) and extended by works on hard state winds (e.g. Mu\~noz-Darias et al. 2019) was put to the test by the luminous outburst of the nearby black hole X-ray binary V404 Cyg in June 2015.

V404 Cyg (this is the optical variable designation, it became also known as the X-ray source Ginga (GS) 2023+338) is a black hole binary which drew widespread scientific attention when it underwent a major X-ray outburst in 1989 (Makino et al. 1989; Kitamoto et al. 1989) which was accompanied by a bright optical and radio outburst (Wagner et al. 1991; Han \& Hjellming 1992).  It subsequently faded back to an anomalously-luminous `quiescent' state (Wagner et al. 1994), where it was possible to track the radial velocity of the companion star around the orbit and place a firm lower limit on the mass of the compact object of $\ga 6 M_{\odot}$ (Casares, Charles \& Naylor 1992), which was one of earliest and strongest direct pieces of evidence for the existence of astrophysical black holes. The system has the second-largest orbit for a black hole X-ray binary after GRS 1915+105 (McClintock \& Remillard 2006), and has a precisely-determined distance of $2.39 \pm 0.14$ kpc obtained via radio parallax measurements (Miller-Jones et al. 2009), confirmed (to within 10\%) by Gaia (Gandhi et al. 2018).

At 18:31:38 UT on June 15, 2015, the orbiting Neil Gehrels Swift Observatory ({\em Swift}) X-ray mission triggered on a burst of gamma-rays from the northern sky. It was soon recognised (Barthelmy et al. 2015) that the burst had come from the direction of V404 Cyg, which triggered a huge multiwavelength campaign on the system. Our radio observations with AMI-LA and later with e-MERLIN, were the earliest and highest cadence radio observations during the outburst; they form the core of this paper and are described in more detail in the next section. Many other notable and important results have been published, including strong cyclic optical oscillations (Kimura et al. 2016), association of positron annihilation signatures with the event (Siegert et al. 2016, but see Motta et al. 2021b), the presence of a strong neutral wind and nebular phase (Mu\~noz-Darias et al. 2016), direct estimation of the magnetic field in the corona/base of the jet (Dallilar et al. 2017), measurement of the size scale of the optical jet base (Gandhi et al. 2017) and very strong and variable X-ray absorption and outflow (King et al. 2015; Motta et al. 2017a,b). Most relevant to this study, Tetarenko et al. (2017, 2019) discuss phases of strong mm and radio flaring, and Miller-Jones et al. (2019) reported mildly-relativistic, milli-arsecond scale radio jets with a strongly varying position angle.

\begin{figure}
\centerline{\epsfig{file=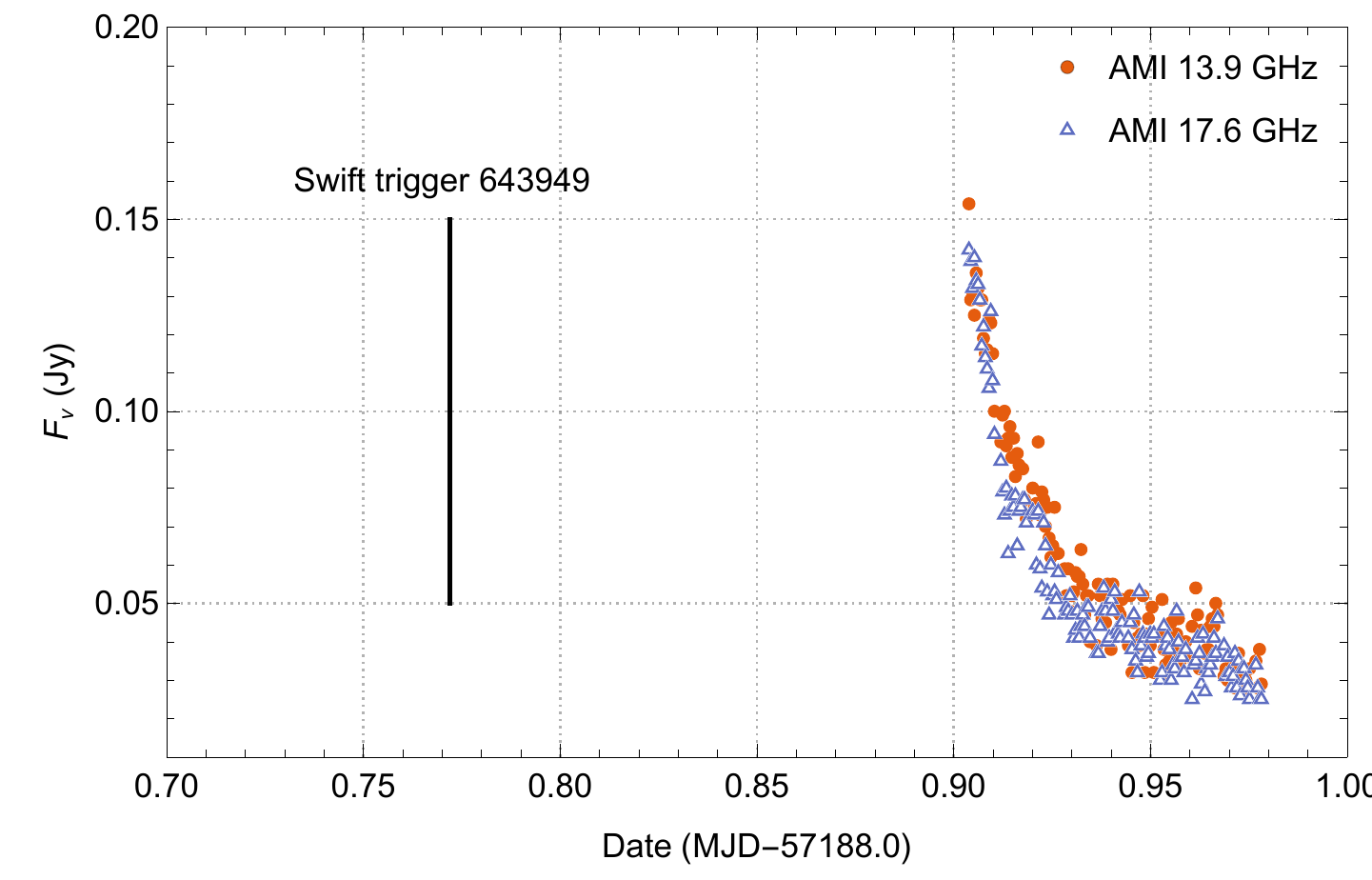, angle=0, width=8cm}}
\caption{AMI-LA observations of V404 Cyg on 2015 June 15. These observations
were triggered and performed automatically, with no human intervention, as part
of the ALARRM programme, and were the earliest radio detections of the 2015
June outburst.}
\label{day1}
\end{figure}

\begin{figure*}
\centerline{\epsfig{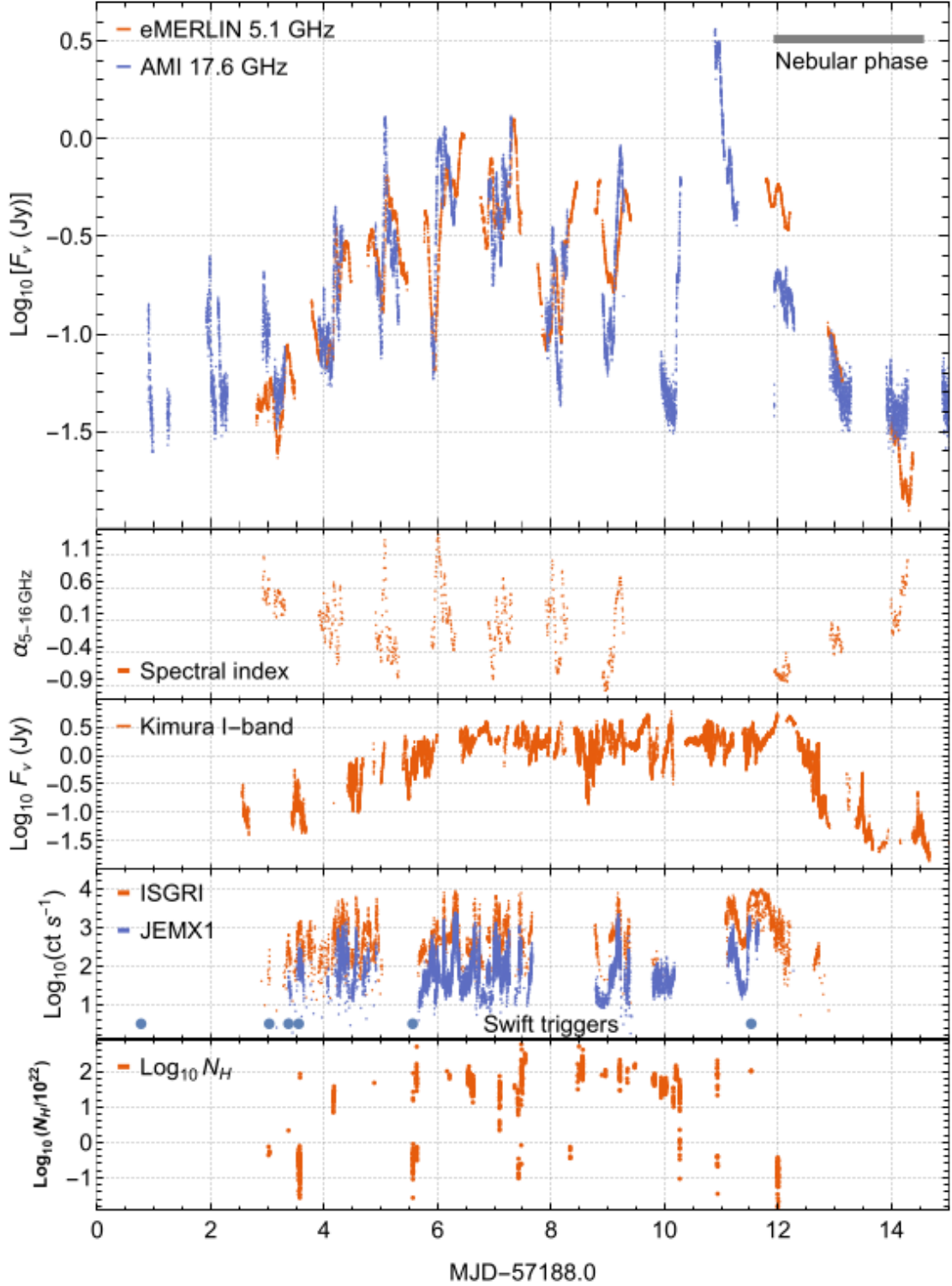}}
\caption{Summary of AMI and e-MERLIN radio observations of V404 Cyg covering the 2015 June outburst. The upper panel (log scale) plots the data from both radio telescopes,illustrating the highly variable (very `flare-y') nature of the radio emission, which reached its peak with one or more very large amplitude ($> 2$ Jy) flares around days 10--12. The grey bar in the upper right indicates the time of the optical nebular phase reported by Munoz-Darias et al. (2016). 
The second panel shows the evolution of the eMERLIN-AMI spectral index (5--16 GHz): the behaviour is qualitatively as expected, optically-thick to -thin transitions superposed on a general trend to steeper indices as more optically thin material accumulates. By day 14 the system seems to have already recovered to a spectral index of $\sim 0$.  The third panel shows the (dereddened) I-band red-optical data from Kimura et al. (2016). The fourth panel shows the JEMX-1 (5-10 keV) and ISGRI (25-200 keV) X-ray data from INTEGRAL (log scale). Like the radio emission, but even more so, the X-ray emission is extremely irregular, composed of many flare-like events rather than the smoother profile typically associated with the middle phases of X-ray transient outbursts. Blue circles towards the bottom of this panel indicate the times of {\em Swift} burst triggers; the first burst triggered the robotic observations presented in Fig 1. Finally, the lower panel shows the very strong variations in neutral hydrogen absorption reported by Motta et al. (2017a,b).}
\label{summary}
\end{figure*}

\section{Radio observations}

\subsection{AMI-LA}

The Arcminute Microkelvin Imager, Large Array (AMI-LA) is a radio telescope operating in Cambridge, UK.
It comprises eight 13m antennas, operates in the 13.9--18.2 GHz frequency range, and has baselines in the range 18-110m. For a full technical description of the telescope see Zwart et al. (2008) and Hickish et al. (2018).

Since 2013 we have been operating a programme on AMI in which it will respond rapidly and robotically,
with no human intervention, to well-localised {\em Swift} gamma-ray triggers (Staley et al. 2013). This programme is referred to
as the AMI-LA Rapid Response Mode, {\em ALARRM}, and is the only regularly-operating robotic radio facility in the
world. The original motivation for the ALARRM mode was to study early-time radio emission from GRBs, and results include several very early-time radio detections as part of a broad unbiased survey (Anderson et al. 2014, 2018). As the programme responds to all {\em Swift} alerts, other science has occasionally resulted from ALARRM, e.g. a prompt radio transient associated with a gamma-ray superflare from a nearby rapidly rotating young red dwarf (Fender et al. 2015). Alongside the ALARRM programme, we perform an extensive set of ad-hoc monitoring and follow-up observations of transients and variables with AMI-LA  (e.g. van Velzen et al. 2016, Anderson et al. 2017, Mooley et al. 2017, Bright et al. 2018, 2020, Motta et al. 2021a).

As part of the ALARRM programme, AMI responded automatically to the {\em Swift} burst alert from V404 Cyg on 2015 June 15. The source was below our $30^{\circ}$ elevation limit
at the moment of trigger but rose to an observable elevation within three hours, at which point AMI-LA began recording data automatically.

These first data from the 2015 June outburst already revealed a radio source over a 1000 times more luminous than the typical quiescent level (Gallo, Fender \& Hynes 2005), and apparently declining from a brighter flare the peak of which we had missed (this very early radio detection was reported in Mooley et al. 2015). This first set of data are presented in Fig 1. Such a nearby luminous black hole outburst is an important and rare event, and we immediately suspended our ALARRM mode and monitoring of most other variables to focus all available observing tine on V404 Cyg. As a result we have extremely extensive radio coverage of the event, which we present in this paper. Previous papers have utilised subsets of these data (e.g. Munoz-Darias et al. 2016a, Shahbaz et al. 2016, Gandhi et al. 2017, Motta et al. 2017b, Dallilar et al. 2018).

The AMI-LA data have been split into three channels, centred at 13.88, 14.63 and 17.63 GHz respectively, and flux densities measured at 40 second time resolution. In most figures we use the upper AMI band in comparison with the eMERLIN data, which is centred at 5.1 GHz, in order to provide the greatest spectral leverage. In some cases, in particular when there is no eMERLIN coverage, we compare the highest and lowest AMI channels; we do not use the middle channel data in any significant way in our analysis (the source is always bright enough to detect in an individual channel). Note that the noise in 40 seconds in each channel is at worst a few mJy, which means that essentially every measurement is a clear detection of the source.

\subsection{eMERLIN}

Once the bright and ongoing outburst of V404 Cyg was apparent, we submitted a Target of Opportunity request to the \textit{e}-MERLIN radio array (Garrington \& Beswick 2016). The proposal was accepted and we acquired extensive coverage of the outburst also at 5 GHz, complementing our AMI-LA observations, commencing on June 17 2015. Daily full-track observations were performed over the following week, using all out-stations and including the Lovell telescope, when available. Three further observations were performed between June 26 - June 28. In all cases, 3C286 was used as a flux calibrator and OQ208 used as a band pass calibrator. The phase calibrator was J2025+3343 and the observations used a target-calibrator cycle time of 8:2 minutes.

Each epoch was calibrated individually in AIPS, following the standard procedures described in the \textit{e}-MERLIN AIPS Data reduction pipeline (Argo 2015). Once all 10 observations had been calibrated, the data were phase-rotated in the \textit{uv}-plane to place V404 Cyg at the phase centre and were then concatenated together into one file. The concatenated file was self-calibrated on the phase calibrator in order to tie all of the complex gains together between the 10 observations. Imaging procedures were also performed in AIPS on subsets of the data where the flux was stable and resolved jet ejecta were not found at any point: in short, V404 Cyg remained unresolved throughout the entire two weeks of observations at \textit{e}-MERLIN scales (typically $\sim 40$ mas resolution).

After calibration and imaging procedures, the concatenated data were loaded into CASA Version 5.6. To produce the light curve, we used the \textsc{casabrowser} tool to extract the visibilities, and average over all baselines and spectral windows per each 60~s scan to create a single data point per time stamp. These data points are shown in Figs 1, 7 \& 8 and derived products in other figures. The flux errors are dominated by flux calibration errors in 3C286, which are of order 5 per cent.

\subsection{Optical and X-ray data}

We use the red-optical I-band data from Kimura et al. (2016). To convert from magnitudes to mJy we use a zero point flux density of 2550 Jy, and we deredden by $A_I = 2.4$ magnitudes.

We furthermore utilise X-ray data from INTEGRAL.
INTEGRAL started observing in on 17 June 2015, during Revolution 1554, and continued to observe until 13 July 2015. Given the importance of V404 Cyg, as a service to the scientific community, the INTEGRAL groups provide a number of data products, including the JEM-X and IBIS-ISGRI light curves used here. The INTEGRAL data can be found at {\bf http://www.isdc.unige.ch/integral/analysis\#QLAsources} (Kuulkers et al. 2015; Kuulkers \& Ferrigno 2015), where more information about the data is available. Part of the data used in this work have been presented in Motta et al. (2017b). 

The IBIS/ISGRI light curves were extracted using OSA 10.2 in the energy bands 25-60 and 60-200 keV adopting a 64~s time bin. The JEM-X light curves were extracted in the energy bands 3-10 and 10-25 keV, with a time bin of with a 8~s. In this work, for each instrument we combined the light curve obtained in the two energy bands into one total light curve per detector, which we show in Figure 2 (second panel from the bottom).

\section{The radio emission from V404 Cyg in 2015 June}

The automatically-triggered observation of V404 Cyg on 2015 June 15 (Fig 1) already revealed the source to be two to three orders of magnitude brighter than its flat-spectrum quiescent level of $\sim 0.4$ mJy (e.g. Gallo, Fender \& Hynes 2005; Plotkin et al. 2019). The radio emission appeared to be in the decay phase of a flare of peak flux density $\geq 150$ mJy, and is consistent with being optically thin in this phase. 

Fig \ref{summary} presents an overview of the radio emission from V404 Cyg during the first 15 days of 2015 June outburst, summarising the AMI channel 3 (17.6 GHz) and e-MERLIN (5 GHz) results, alongside the radio spectral index, I-band optical data and 5--200 keV X-ray monitoring from the INTEGRAL spacecraft. Although the focus of this paper is on the radio emission and its use as a tracer of particle acceleration and kinetic feedback, it is striking that the X-ray emission (note the logarithmic scale) is characterised by repeated flaring by a factor of $\geq 100$, which is rather unlike `normal' X-ray transient outbursts (e.g. Dunn et al 2011). As will be quantified later, in terms of cadence and luminosity space, this is the most comprehensive radio monitoring ever performed of a black hole X-ray binary outburst.


Like the X-ray emission, the radio emission is also characterised by rapid and repeated flaring and associated spectral index changes (Fig 2, second panel). We define this spectral index $\alpha$ here in the sense that the flux density varies with frequency as $S_{\nu} \propto \nu^{\alpha}$. Over the first ten days of activity, the typical synchrotron flare spectral index variations (optically thick on the rise, thin on the decay) are superposed on a general trend towards more negative spectra. This presumably arises from an increasing contribution of optically thin emission from preceding flares. The major radio flare on/around day 11 (MJD 57199.0) is the apex of this trend and following it the spectrum slowly recovers to its approximately flat ($0 \leq \alpha \leq 0.5$) pre-flare spectrum over a timescale of about 4 days (this spectrum is consistent with that measured in quiescence: Gallo, Fender \& Hynes 2005; Rana et al. 2016). We note that although the major activity was confined to a $\sim 15$-day period there is a significant radio flare nearly 20 days after the initial trigger, and there was further radio flaring during the smaller subsequent outburst in December 2015 (Munoz Darias et al. 2017).

\subsection{Distribution of radio fluxes and a flux-rms relation}

\begin{figure}
\centerline{\epsfig{file=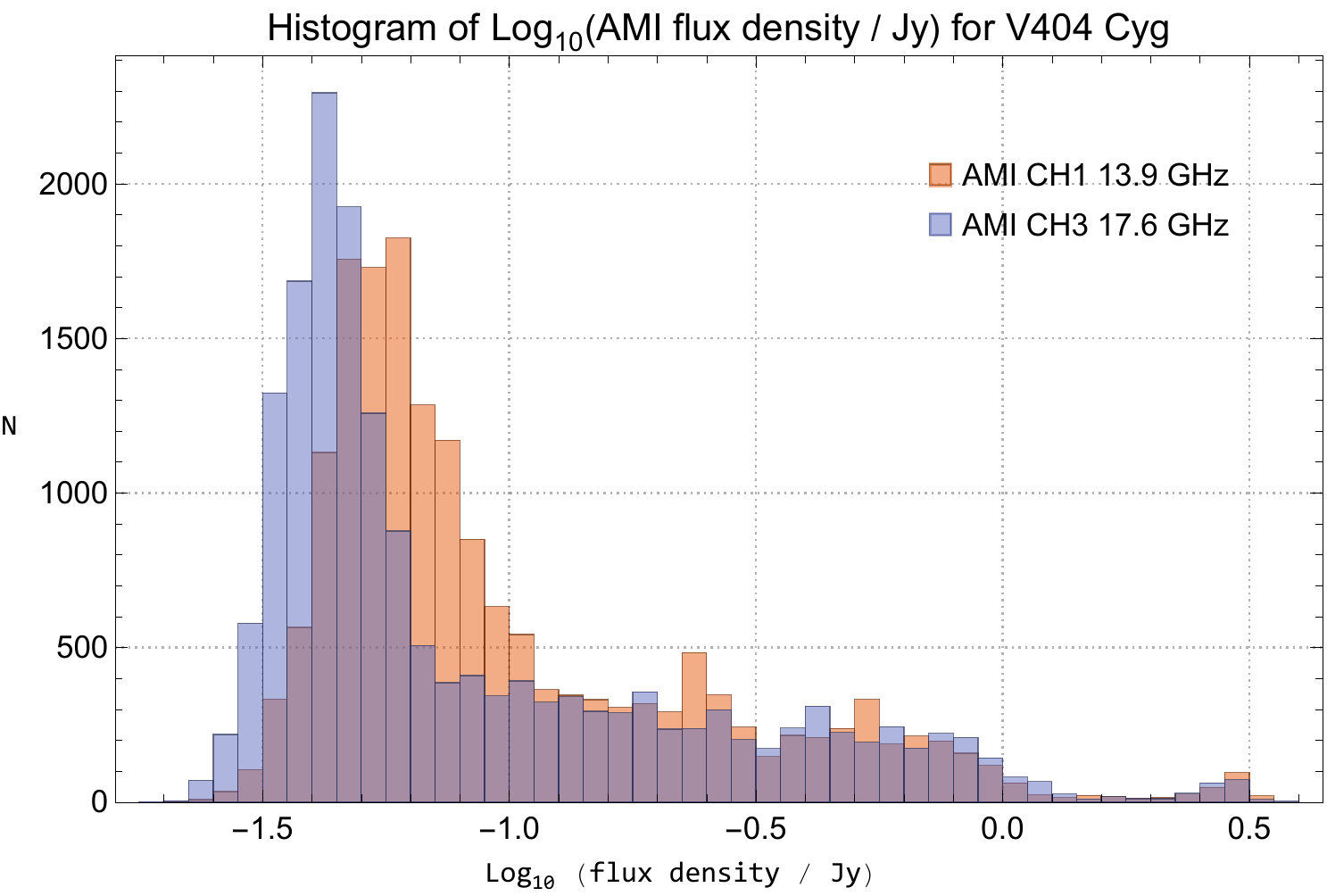, angle=0, width=8cm}}
\caption{Distribution of the Log$_{10}$ of the AMI flux densities, in channels 1 and 3. Both distributions show a broad peak at around 40-50 mJy plus a long tail to higher flux densities. The CH1 peak is shifted to slightly higher flux densities. See main text for more discussion.}
\label{h13}
\end{figure}

Counting each 40-sec integration, we have 17 658 AMI radio flux measurements of V404 Cyg.
Fig \ref{h13} presents the distribution of the logarithm of these for the 15 main days of the outburst. The distribution appears to be approximately Gaussian-like at low flux densities with a long tail to higher flux densities. The low-flux density peak is shifted to higher values in Channel 1 (13.9 GHz), where it is at around 50 mJy, compared to Channel 3 (17.6) GHz, where it is around 40 mJy. Ignoring simultaneity, for now, this corresponds naively to a mean spectral index in this part of the histogram, within the AMI band, of around $-1$. 

In order to probe the long term variability in the radio data, we generated the Lomb-Scargle periodogram (Lomb 1976; Scargle 1989) for the entire AMI dataset, covering almost 5 orders of magnitude in frequency ($\sim$10$^{-7}$-10$^{-3}$Hz). We used the "standard" Lomb-Scargle normalisation, where the power is divided twice by the variance of the signal. In Figure 4 we show the  periodogram from the data (orange dots), and that from the window function (blue dots), which allows one to easily distinguish what features are real, and what are to be ascribed to the data sampling, such as the two large spikes at 0.0015 and 0.003 Hz (i.e. 600 and 300s). The spike at 0.0015~Hz visible in both the data and window function, corresponds to the length of the on-target scans in the AMI observations (600s), which are interleaved by phase-calibrator scans (100s), while the spike at 0.003~Hz is a strong harmonic to the same periodic feature. The peak at $\sim$10$^5$s visible in the window function is related to the time-scale set by the average length of the AMI daily runs, i.e. which lasted approximately 9 hours. At frequencies below $\sim$10$^{-3}$~Hz there is significant power in excess of the window function. A similar analysis of four other X-ray binaries, albeit with far smaller data sets, has been presented in Nipoti, Blundell \& Binney (2005).

\begin{figure}
\centerline{\epsfig{file=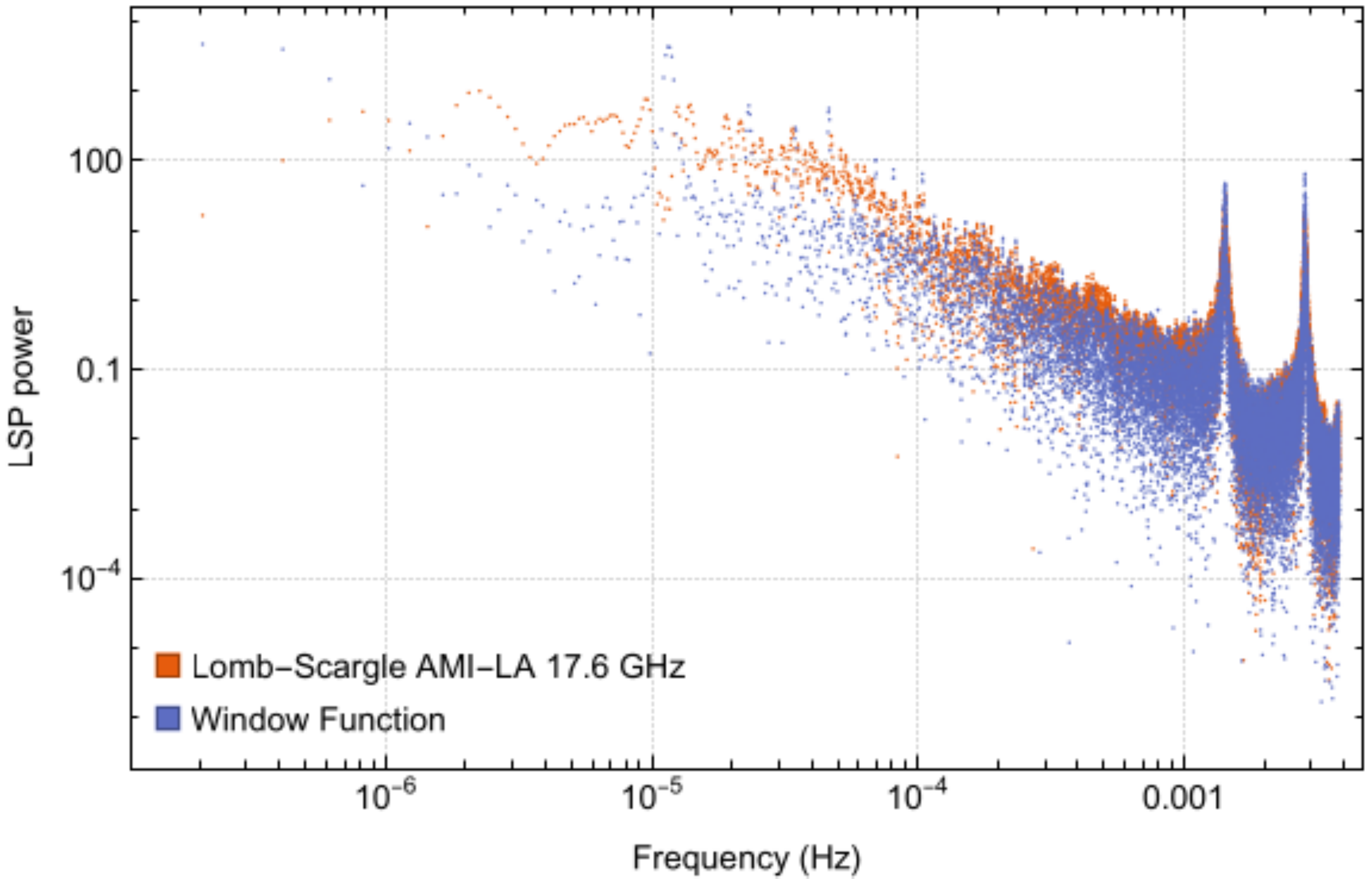, angle=0, width=8cm}}
\caption{Lomb-Scargle periodogram of the 17.6 GHz AMI-LA CH3 data. The LSP power is `standard' Lomb-Scargle normalisation, where the power is divided twice by the variance of the signal.  At frequencies below about $10^{-3}$ Hz there is significant power in excess of the window function.}
\label{LSP}
\end{figure}

\begin{figure}
\epsfig{file=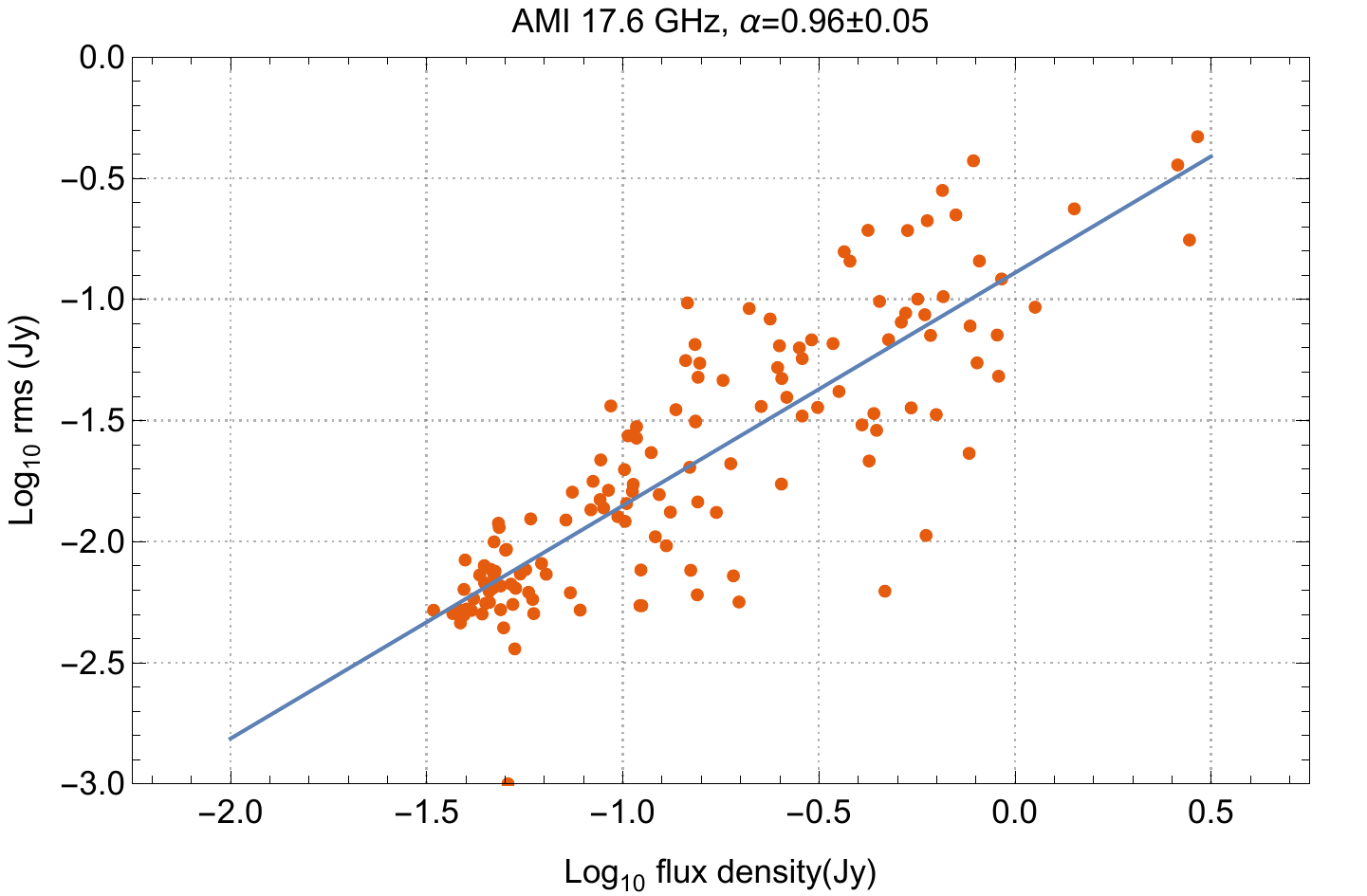, width=8.8cm}\\
\epsfig{file=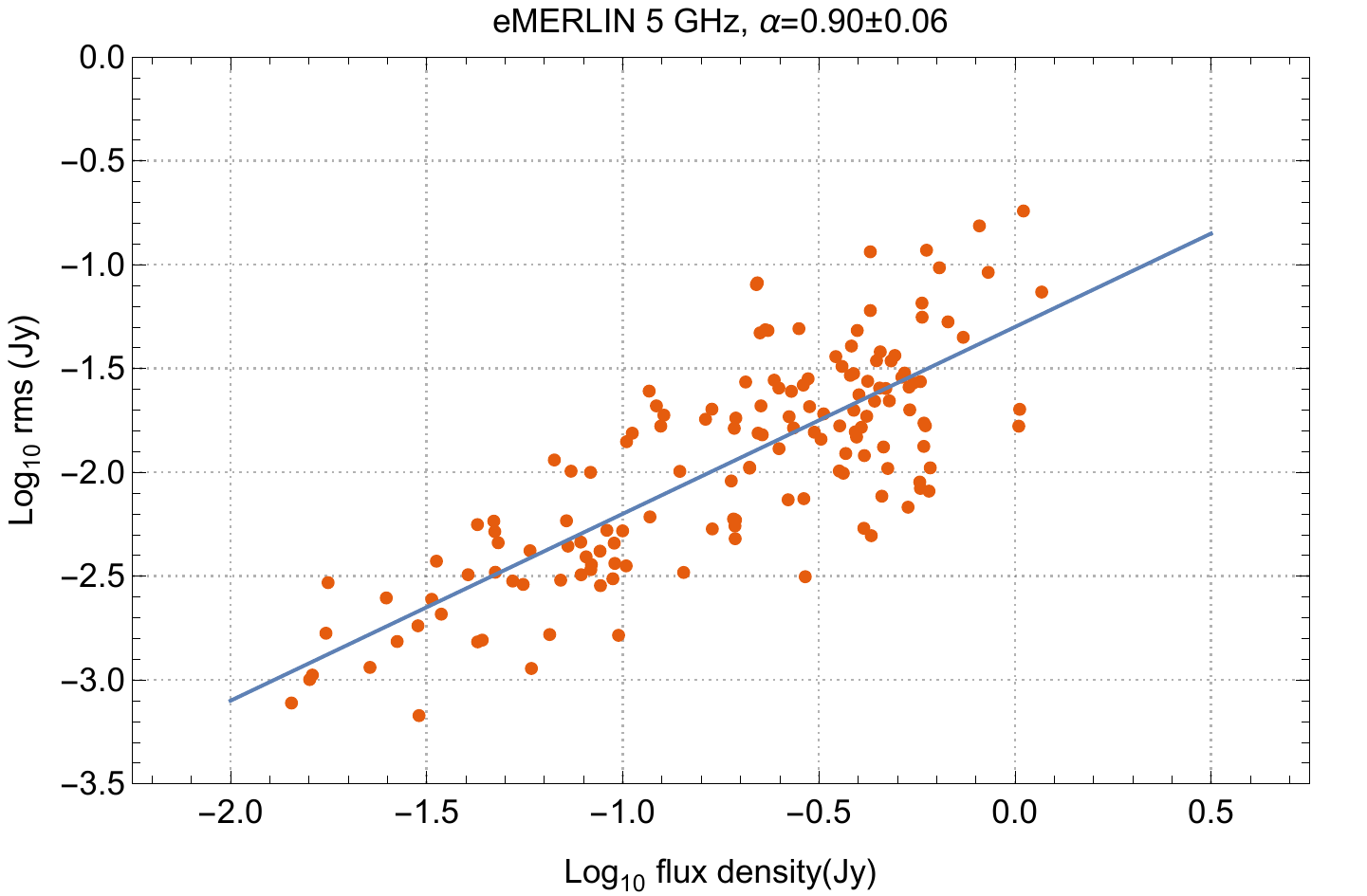, width=8.8cm}\\
\epsfig{file=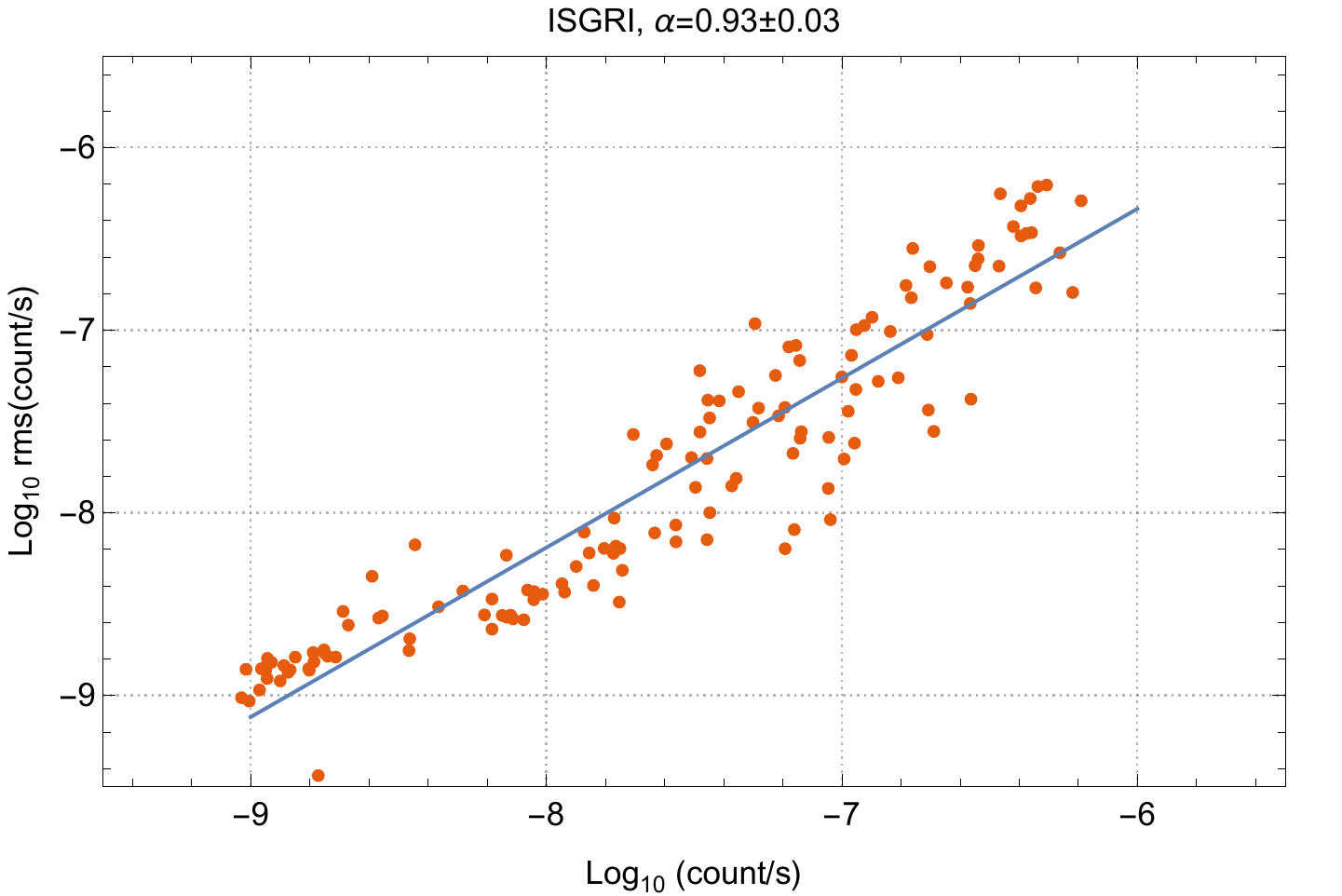, width=8.8cm}
\caption{Flux-rms relation for AMI-LA (highest frequency channel), eMERLIN and ISGRI data sets, binned on one hour timescales. The correlations are very close to linear for all three wavelengths. The AMI and eMERLIN correlations are the first reported flux-rms relation for radio emission from a relativistic jet.}
\label{flux-rms}
\end{figure}

\begin{figure}
\centerline{\epsfig{file=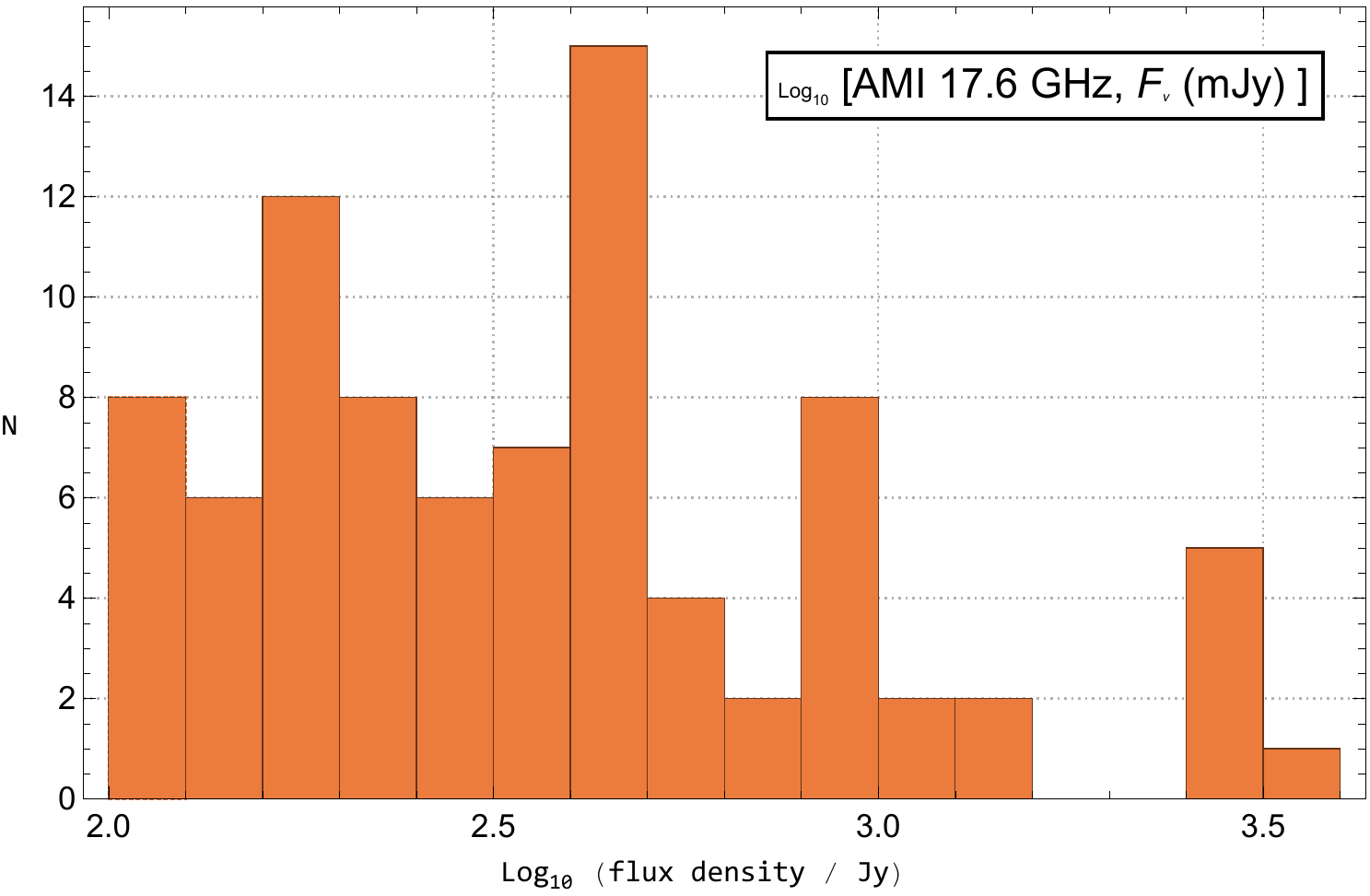, angle=0, width=7cm}}
\centerline{\epsfig{file=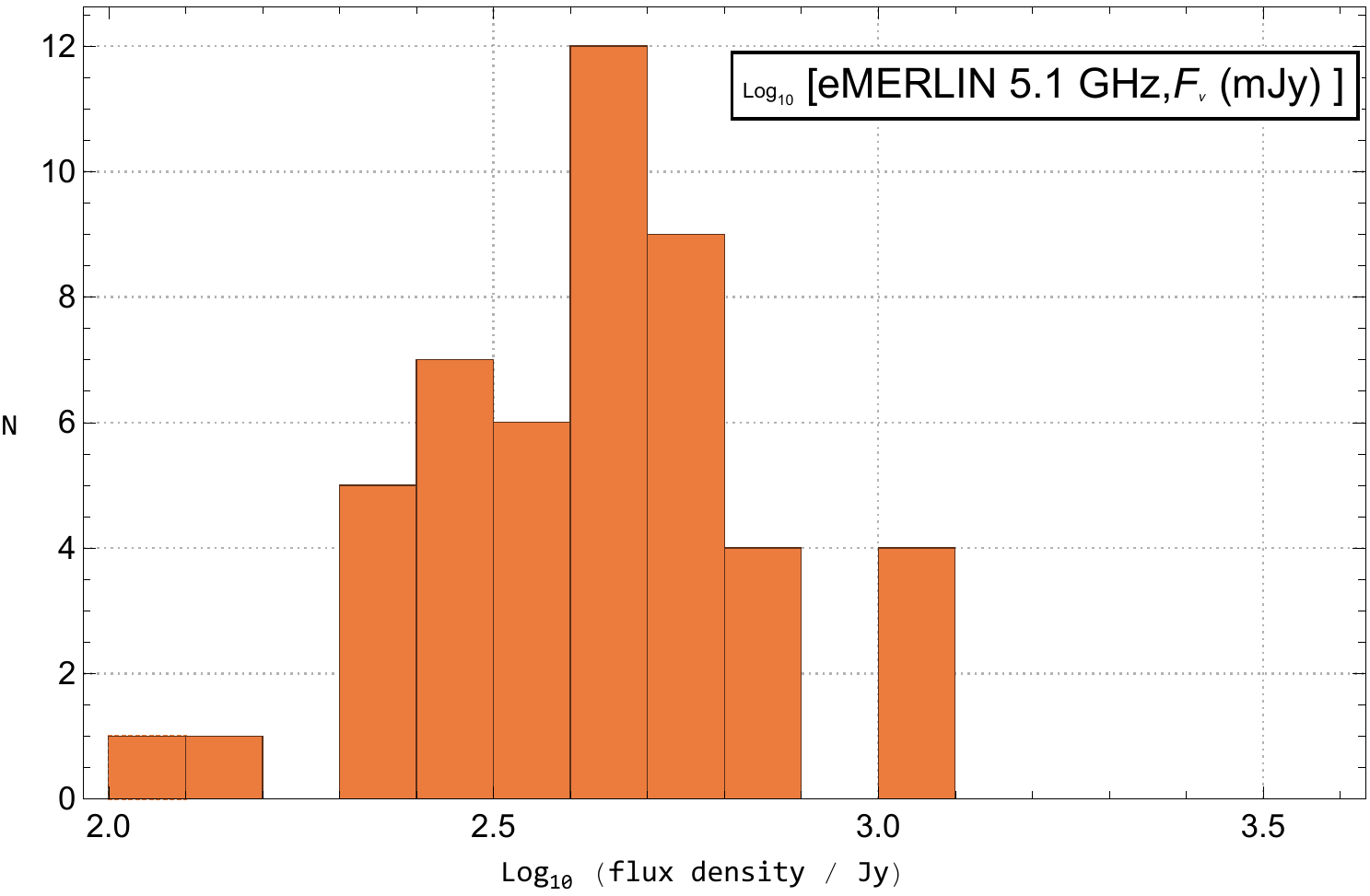, angle=0, width=7cm}}
\caption{Distribution of flare peaks at 17.6 (upper panel) and 5.1 (lower panel) GHz in the AMI and eMERLIN data respectively. See text for details.}
\label{flaredist}
\end{figure}

\begin{figure*}
\epsfig{file=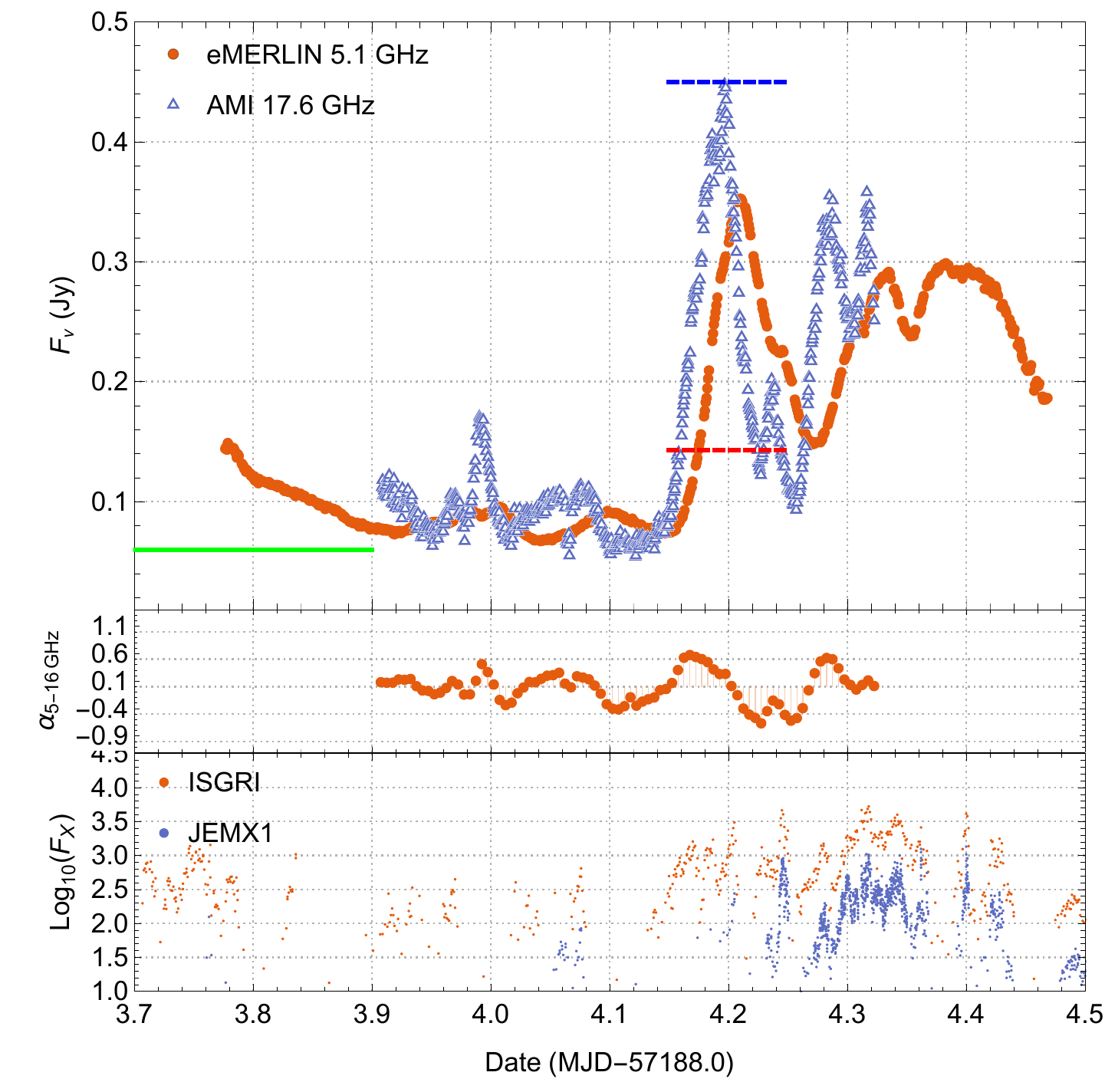, angle=0, width=7cm}\quad\epsfig{file=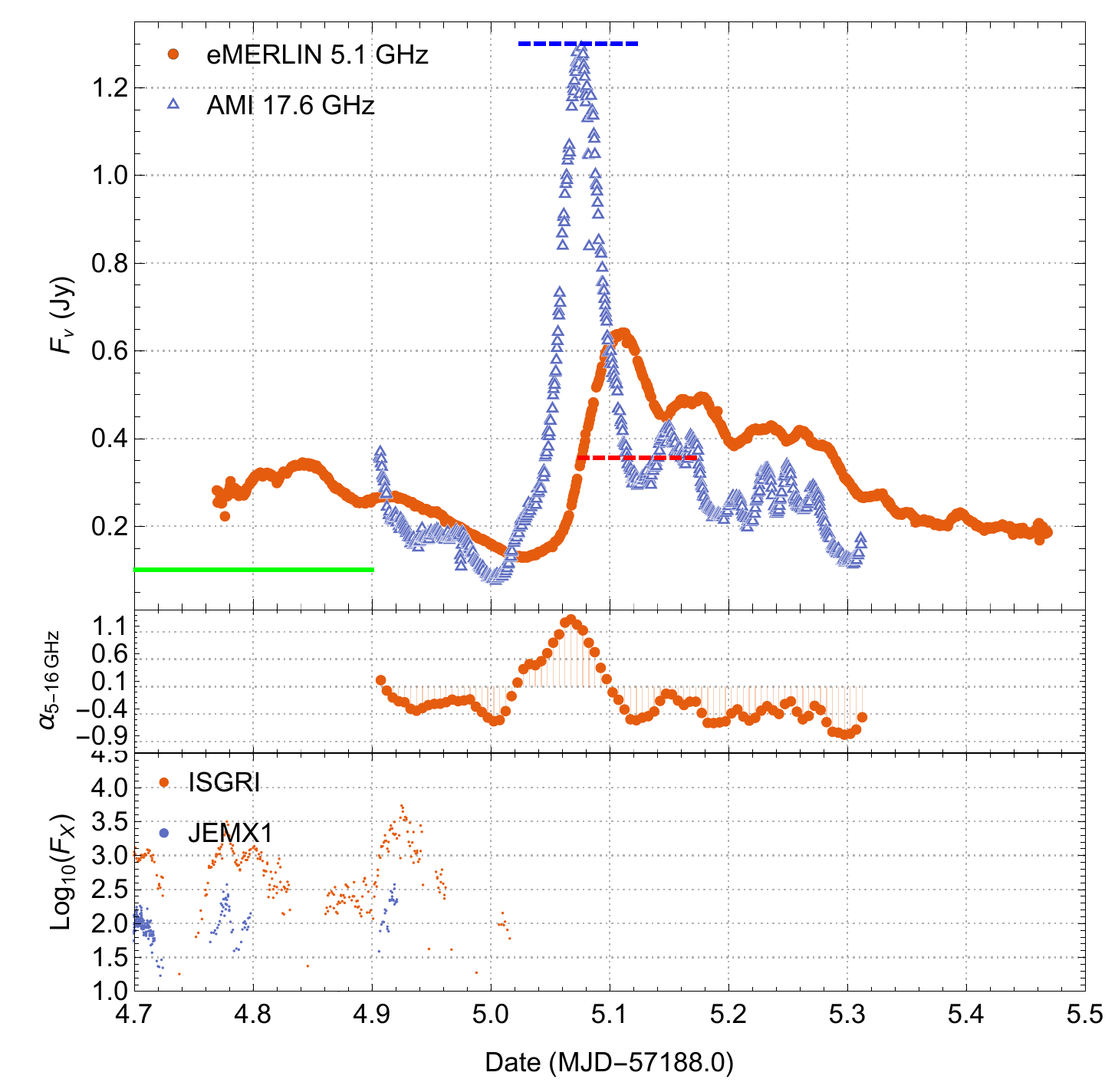, angle=0, width=7cm}\
\epsfig{file=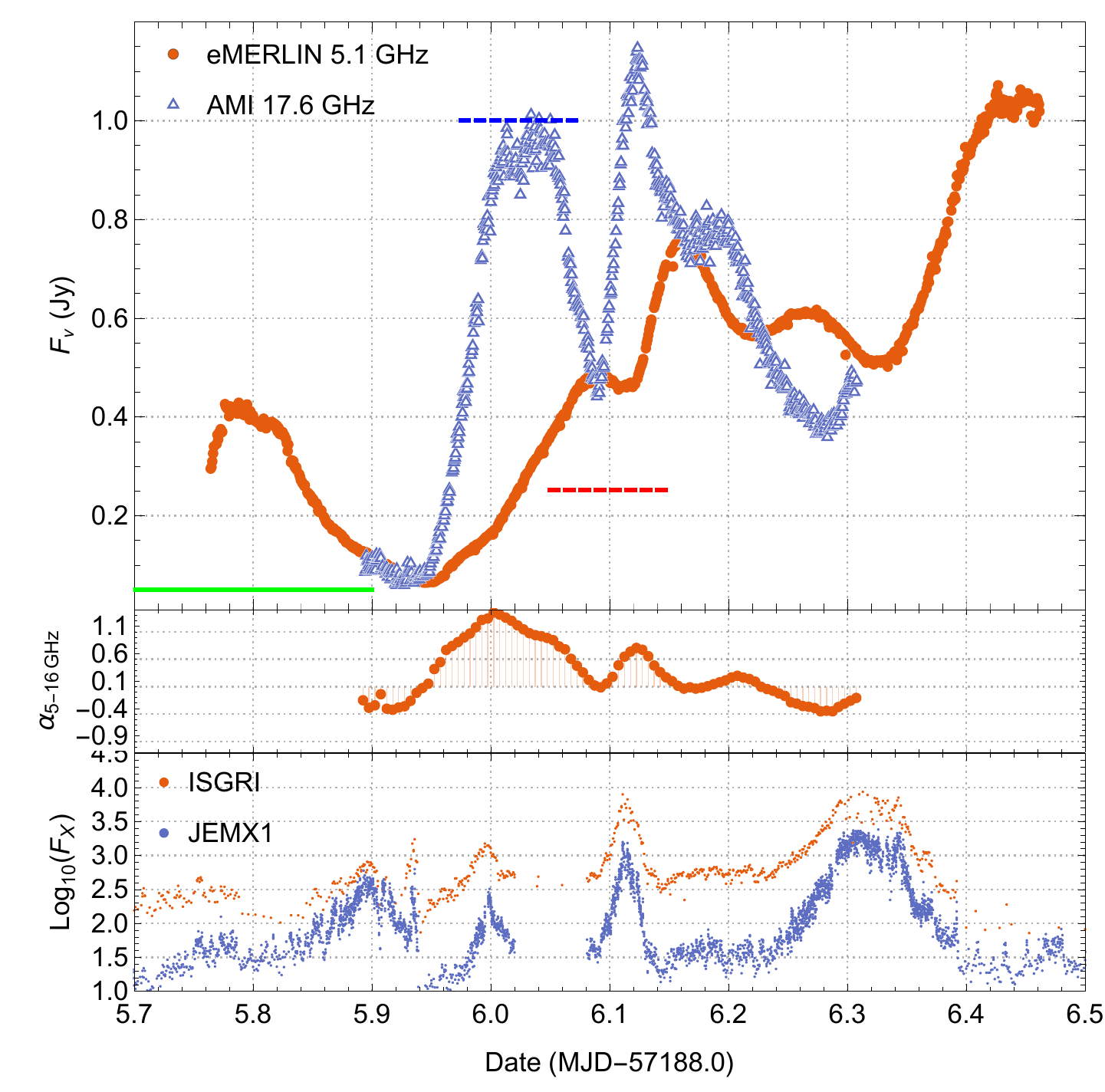, angle=0, width=7cm}\quad\epsfig{file=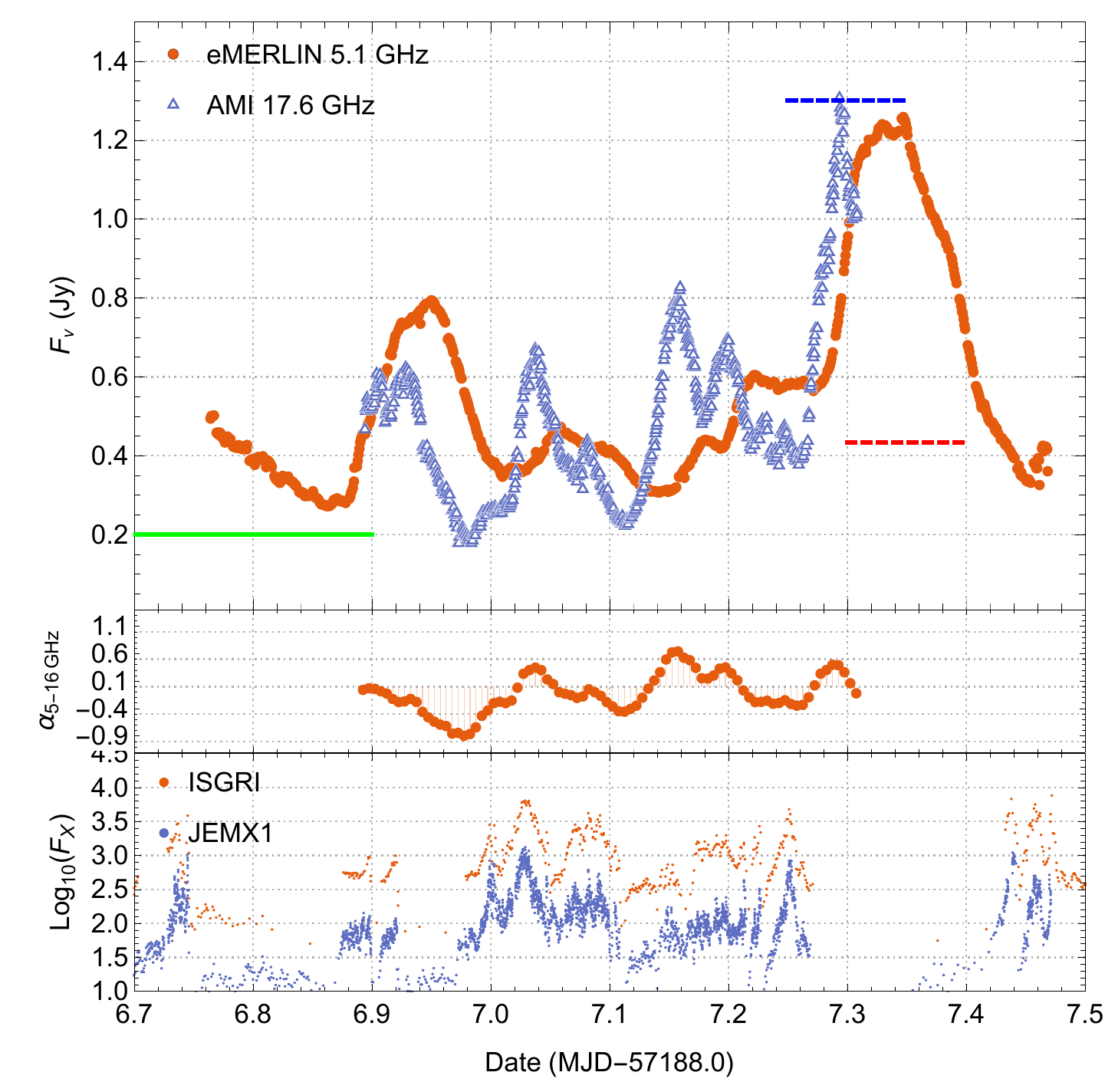, angle=0, width=7cm}\
\epsfig{file=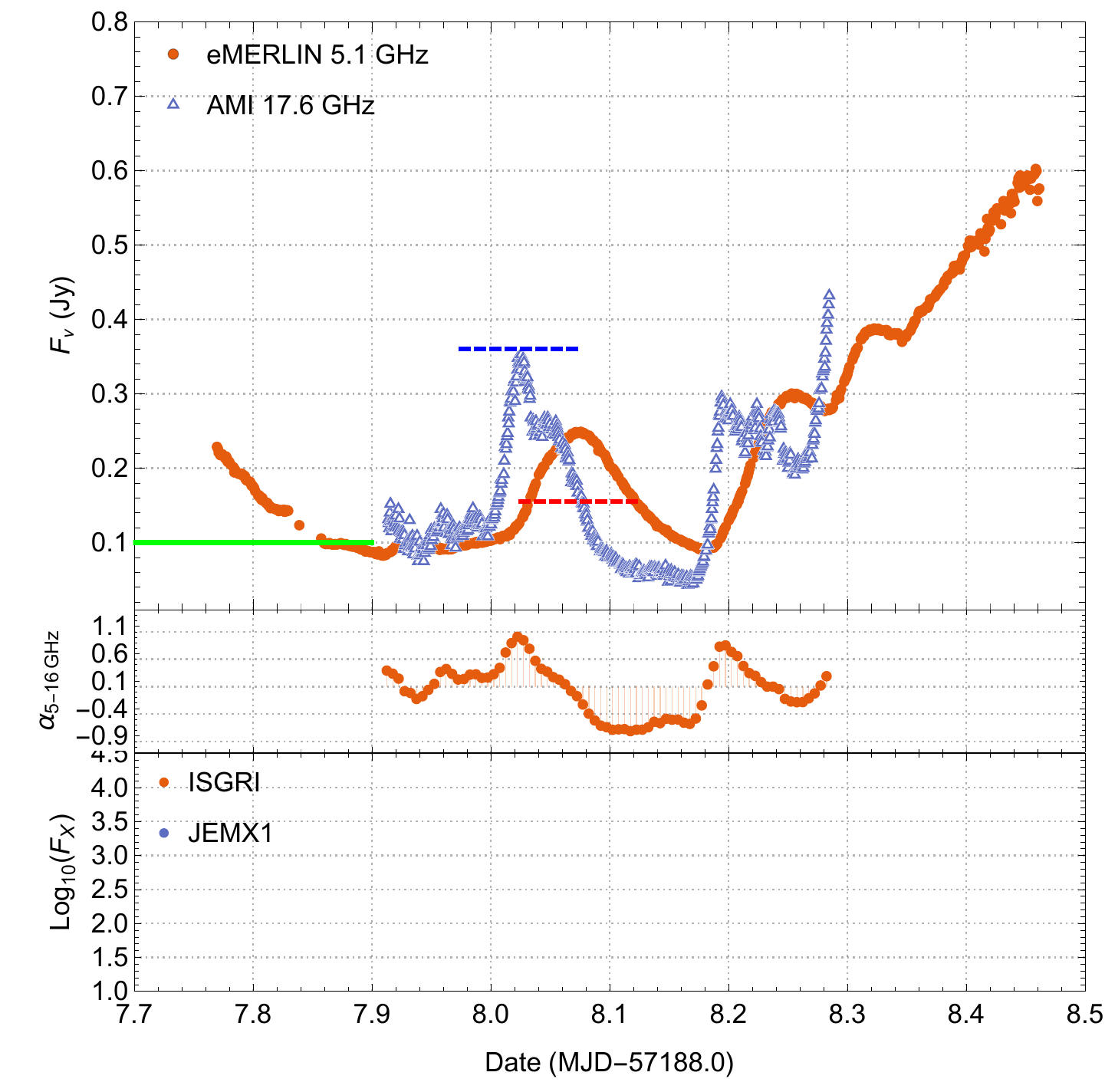, angle=0, width=7cm}\quad\epsfig{file=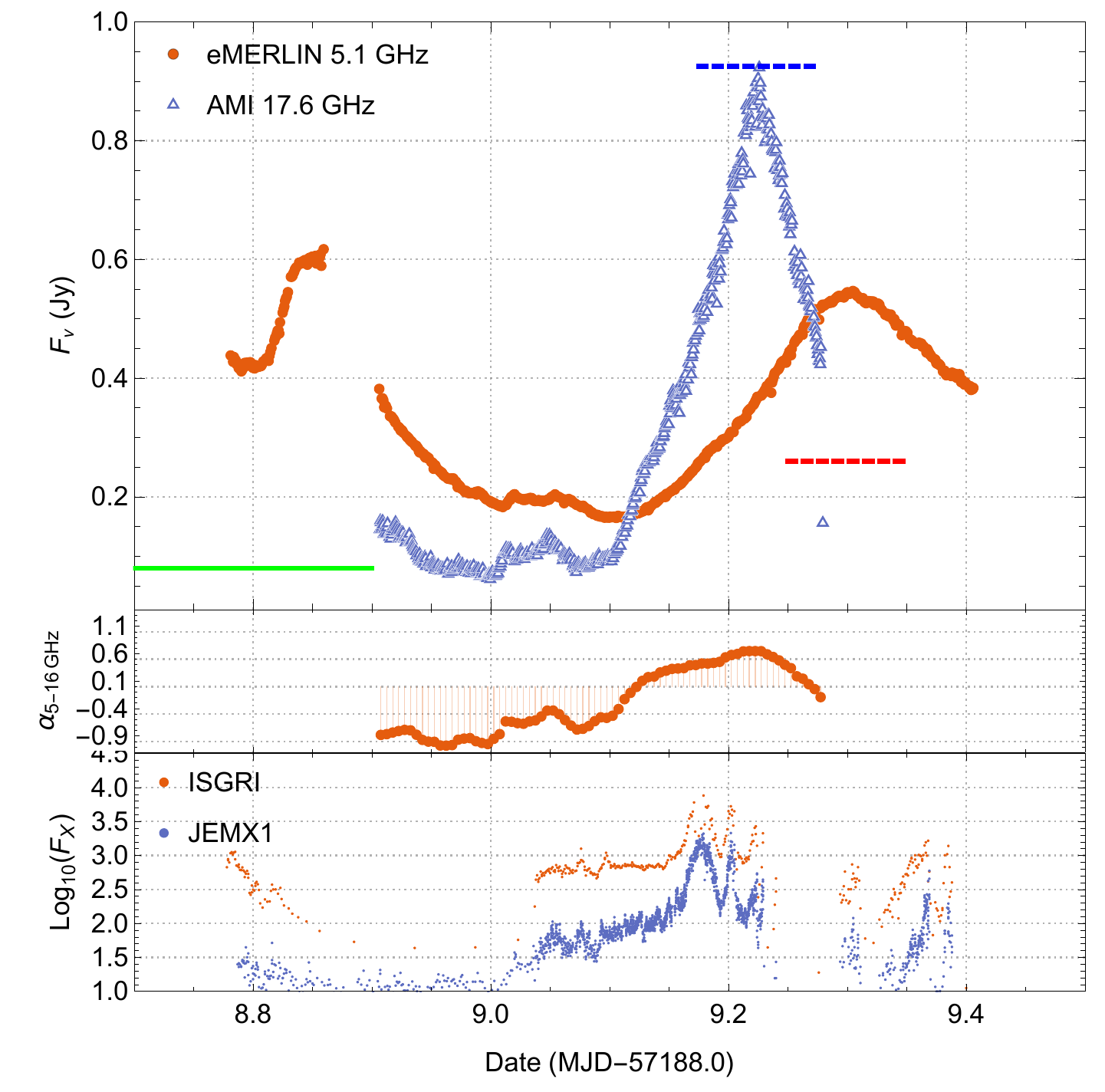, angle=0, width=7cm}\
\caption{Six observing runs in which there was dense coverage with both AMI-LA and eMERLIN, and during which there was significant activity. The upper panels of each figure show radio flux density measured with the two telescope arrays as a function of time. Note that the time interval of each panel is exactly the same, at 0.8 days, whereas the vertical axes have been scaled to match the range of flux densities observed. Three horizontal bars are drawn on each panel; from left to right these are: (green, solid) estimated baseline flux before selected flare event, (blue, dashed) estimated peak flux density at 17.6 GHz for the selected flare, (red, dashed) peak flux density expected at 5.1 based on the van der Laan model taking into account the baseline flux and assuming $p=2.2$ for the electron energy distribution. In every case the lower frequency peak is much stronger than expected for this simple impulsive energy injection model.
The middle panel shows the spectral index $\alpha$ (where $S_{\nu} \propto \nu^{\alpha}$) between the two bands, calculated by rebinning the two light curves to matching intervals of 0.05 days. The lower panel shows simultaneous X-ray observations from INTEGRAL, where available.}
\label{ami-mer}
\end{figure*}

The fact that most of the radio flux densities are distributed in a log-normal way motivated us to look for a flux-rms relationship in the data.
A linear correlation between X-ray flux and root mean square variability (rms) is an established characteristic of accretion flows around compact objects, and provides clues as to signal propagation within accretion discs, although its presence in isolation does not establish an accretion origin
(e.g. Uttley \& McHardy 2001; Uttley, McHardy \& Vaughan 2005, 2017). This is not something which has been investigated in the radio band previously, but the data volume and quality here allows us to test this. In Fig \ref{flux-rms} we plot the rms (measured directly from the light curve, not from power spectra) as a function of flux for the ISGRI X-ray data and flux densities for the AMI and eMERLIN radio data, binned on one hour timescales. We see a clear, near-linear correlation between the flux and rms in both radio bands - this is the first report of the flux(-density)-rms relation for radio synchrotron emission in a relativistic jet. The flux-rms correlation is also seen in the ISGRI data. We use only the first 15 days of the outburst for this analysis, as this covers the vast majority of the flares. The slope of this correlation varies somewhat with averaging interval, but is close to linear for averaging intervals around one hour, which is similar to the typical timescale for a flare event (further discussion below). Interestingly, when averaged on the same timescale, no flux-rms relation is present in the I band optical data.

Interpeting these results is not straightforward, however, for a number of reasons. Firstly, the dynamic range is much larger, and the timescales much longer, than the flux-rms relations studied in X-rays (which come from the inner disc), so it is not clear that they share a common origin. Secondly, the non-LogNormal high flux density distribution in Fig 3 will clearly be contributing to the flux-rms relation (see discussion in Uttley et al. 2005). Therefore, while these relations clearly tell us something significant about the variability patterns of the radio emission right now, and we can be confident that fluctuations from the accretion flow will propagate in some way into the jet (although there is rather limited work in this area, but see Malzac 2013), quantitative intepretation of these results will take further effort beyond the scope of this paper.

\subsection{Distribution of flares}

As noted above, this radio outburst is characterised by a large number of flares of varying amplitude.
We use the find\_peaks function from the scipy.signal module to locate peaks in the AMI data both for the entire data set, and individually for the six days discussed in section 3.3 below, see Appendix for more details. We find 86 peaks in total at 17.6 GHz over the first 15 days of the outburst (i.e. the coverage in Fig 2). For the observing runs centred approximately at days 4.1, 5.1, 6.1, 7.1, 8.1 and 9.1 we find 11, 10, 11, 10, 6 and 5 peaks respectively above a threshold of 100 mJy. This general trend supports the by-eye impression that $10 \pm 1$ days of intense flaring was followed by a brief drop in activity and finally one or more very large flares. Assuming a mean rate of 10 flares of peak height $> 100$ mJy per 0.4 days (the length of an AMI observing run) for the first ten days, this converts to approximately 250 such flares during this period. The distribution of the peak flux density of the 86 peaks identified in the first 15 days of the outburst is presented in the top panel of Fig \ref{flaredist}. The sample is not really large enough to infer any particular statistical distribution, but the large-amplitude flares which occur at the end of the outburst clearly stand out. We repeated the analysis, with the same parameters, on the eMERLIN 5.1 GHz data (Fig \ref{flaredist}, lower panel)  and find a total of 49 individually determined peaks over the 15 days of the main outburst (the lower-frequency peaks are both lower and, usually, broader, which means fewer are automatically identified). The distributions of flares are presented in Fig \ref{flaredist}, where the lower mean flux density of the 5.1 GHz peaks is clear (note also that major flare around day 11 was missed by eMERLIN, which is why the highest peaks are not there at 5.1 GHz).  Later in the paper we shall use these flare measurements, under the assumption that the peaks are due to synchrotron self-absorption, to estimate the kinetic feedback in the radio jet.

\subsection{Individual flares, spectral evolution}

\begin{figure*}
\centerline{\epsfig{file=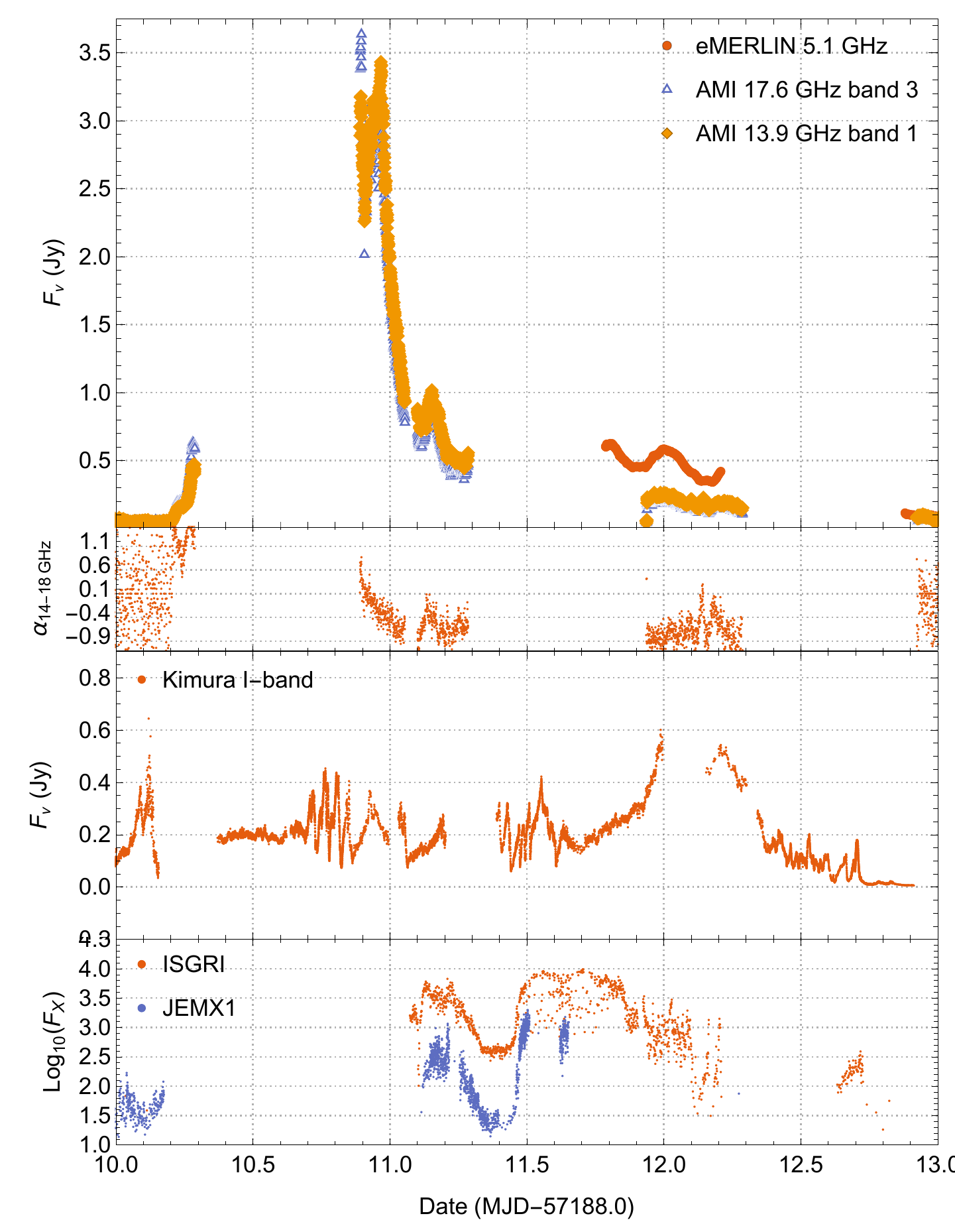, angle=0, width=16cm}}
\caption{The three days around the brightest radio flare, towards the end of the outburst, comparing radio, optical and X-ray emission.
Note that the $>3$ Jy radio flares follow a phase of strong optical oscillations (Kimura et al. 2015). The main flare peak is only captured by AMI, but is bright enough that we can utilise the AMI in-band spectral index, and see clear evolution from optically thick to thin emission following the first, brightest, peak.
The radio flare around day 10 appears to be optically thin in both the rise and decay phases, and probably represents particle acceleration in an extended low optical depth environment (probably a jet-ISM interaction).
}
\label{flare}
\end{figure*}

Given the very high quality of our dual-frequency radio coverage we are able to study a number of flares closely to study their temporal and spectral evolution. In Fig \ref{ami-mer} we present a more detailed view of six sets of observations of flaring periods of V404 Cyg, plotting the radio flux densities from both AMI-LA and eMERLIN, the spectral index between the bands, and the X-ray count rates from JEM-X1 (5 -- 10 keV) and ISGRI (25 -- 200 keV) onboard INTEGRAL. Fig \ref{flare} highlights three days of activity around the largest radio flares which occurred towards the end of the main phase of the outburst, and includes the red-optical I-band flux densities from Kimura et al. (2016).

\subsubsection{Temporal and spectral evolution of flares}

It is clear that the radio lightcurve is largely, possibly entirely, composed of flare events with a range of amplitudes (see later discussion). The behaviour of all of the flares (with the exception of the event on day 10, see below) is qualitatively similar, and as expected for simple models of synchrotron ejecta, in that the higher frequencies peak earlier and stronger (see e.g. van der Laan 1966). This represents very strong evidence that the peaks correspond to the transition from optically thick to optically thin emission. The van der Laan model is often applied to X-ray binary outbursts, and simply considers the evolution of a spherical volume of synchrotron-emitting plasmas as it expands, arising from a single impulsive energy injection event. The trend of the spectral index response is as expected for such events, switching from self-absorbed ($\alpha > 0$) to optically thin ($\alpha <0$) through the peak of the event. This results in a light curve with qualitatively similar shape at two radio wavelengths, but with the lower frequency delayed with respect to the higher. This in turn leads to loops when we plot the simultaneous AMI and eMERLIN data in the radio-radio plane (Figure \ref{loops}). Inspection of this plane reveals that at flux densities above $\sim 100$ mJy the radio emission traces out clockwise loops oscillating between spectral indices of $\pm 1$. The radio spectrum never steepens to the canonical $\alpha = +2.5$ expected for a single optically thick synchrotron component (this may be explicable due to the radio emission at all times being a superposition of components with a range of optical depths, implying that at no time does a {\em single} optically thick component dominate).

When investigated in more detail, however, we see that the flaring activity cannot be explained by the simple van der Laan model. The ratio of peak flux densities at two frequencies in the van der Laan (1966) model is given by

\[
R = \left(\frac{\nu_2}{\nu_1} \right)^{\beta}
\]

where

\[
\beta = \frac{7p+3}{4p+6} 
\]

Comparing the AMI channel 3 (17.6 GHz) and eMERLIN (5.1 GHz) bands, and assuming $p=2.2$ (borne out by observations of optically thin phases in this and other X-ray binaries, although we do note that Tetarenko et al. 2017 derive a wider range in $p$ from their modelling), we expect a ratio of peak flux densities from AMI to eMERLIN of 4.7. The {\em measured} flux ratios between the two bands are in the range 1 -- 2.5, much flatter than can be explained by this simple model. In Fig \ref{ami-mer} we use three horizontal bars to indicate the estimated baseline flux density (green solid line), measured high-frequency flux density (blue, upper dashed line) and the expected flux density at the lower frequency based on the van der Laan model and taking the baseline flux into account (red, lower dashed line). While we acknowledge that there is certainly superposition of multiple events occurring, we have attempted to choose flares which dominate the emission for a short period of time. In every case the lower frequency (eMERLIN) emission is much stronger than expected.  

The brightness temperature is also worth bearing in mind, when considering these flares. For a source at 2.4 kpc and observations at 16 GHz,

\[
T_b \geq 2 \times 10^{15} \Delta S_{\nu} \Delta t^{-2}
\]

\noindent
K, where $\Delta S_{\nu}$ is the variability amplitude in mJy on a timescale $\Delta t$ seconds. This assumes that the source size $r=c \Delta t$; for a smaller size the brightness temperature will be larger.
For example, for the recently-inferred physical expansions speeds of X-ray binary ejecta (Tetarenko et al. 2017; Fender \& Bright 2019) of $\sim 0.1c$, the brightness temperature would be 100 times larger.
Setting $T_b \leq 10^{12}$K we can rearrange and see that

\[
\Delta t \geq \sqrt{2 \times 10^3 \Delta S_{\nu}}
\]

\noindent
(where the constant in the above equation is $2 \times 10^5$ for expansion speeds of $0.1c$). Therefore a 5 mJy flux change should take $\geq 100$s and timescales scale as $\sqrt{\Delta S_{\nu}}$. 

In inspecting individual flare events, we noticed that there appeared to be quite different rise rates between events of similar amplitudes. In Fig \ref{rise1} we show a collection of eight flare events which we have directly overlaid, in each case selecting by eye a time and baseline flux (as indicated in the figure). We note that three events appear to rise at a rate which is about three times as fast as the other five. While this sample is small, it is suggestive of a dichotomy in behaviour. Furthermore, we have over-plotted onto these events two lines representing $F_{\nu} \propto \Delta T^3$ and $F_{\nu} \propto \Delta T$. Neither of these lines is a fit, they were simply chosen to start at (0,0) and cross at (0.03,0.8), near to the peak of the brightest event. The cubic relation is that derived by van der Laan (1966) for a simple expanding synchrotron-emitting blob, whereas what we see is that the brighter events in fact appear to rise close to linearly with time (these distinctions become blurred if time zero for the flare was actually significantly earlier than the first observed indication of flux increase). Examining these same events over a longer timescale, as presented in Fig \ref{rise1} (lower panel) we see that the faster-rising events are all rather isolated events, and the slower-rising events are all part of larger complexes of radio flares which show plateaux and changes in rise rate.

The strongest radio flare observed by us during the outburst was that of June 26 (Fig \ref{flare}) which peaked at over 3 Jy in the AMI band. Unfortunately there is no direct coverage of the event with eMERLIN (although observations did resume the following day), but we can inspect the spectral evolution using the highest and lowest AMI bands (second panel, Fig \ref{flare}). Here we see that the first and highest observed flare peak appears to be optically thick, but that the spectral index is decreasing rapidly following this peak and within $\leq 0.2 {\rm d} \sim 5 {\rm hr}$ the emission is optically thin. In fact the second peak recorded is brighter in the lower frequency band. A small flare detected during the decline provides a temporary flattening of the spectrum, and then finally around day 12 we see an event which appears to be optically thin through its rise and decay. This sequence of events implies a powerful optically thick ejection which led, within one day, to an optically thin event away from the inner regions of the jet, and that between these two events further compact ejections were taking place.
Also noteworthy are the strong optical oscillations before/during the strongest radio flaring: although they are not unique, and we do not have simultaneous radio coverage, they do appear to increase in amplitude as the radio peak is approached.

\subsubsection{The optically thin flare}

Peaking around day 12.0, we see a relatively slow flare event which appears to remain optically thin throughout (note that this is supported both by the eMERLIN-AMI spectral index [Fig 1] and also the in-band AMI spectra index [Fig 8].). Unlike the optically thick flares (every other flare as far as we can tell), the rise phase ($\Delta t \sim 0.1$d) for this event cannot correspond to an evolution from optically thick to optically thin, with the peak occurring around optical depth unity. Instead, the rise phase must correspond to the timescale for the maximum (peak optically thin flux) electron population to be accelerated and/or for the associated radiation to reach the observer.

Such optically thin flares are rare in X-ray binaries but have been seen before to follow sequences of powerful optically thick flares, such as the case for flare 'V' from Cygnus X-3 in 1994 (Fender et al. 1997). Noting that this one strong optically thin flare from V404 Cyg occurs one day after the largest (optically thick) flare(s) from the source, we can also consider whether there might be some connection between them. In the following we make use of minimum energy 'equipartition' analysis both in the general case for an optically thin case where the size may be estimated (see e.g. Longair 1994) or the special case of emission at optical depth unity (i.e. the peak of an optically thick flare) as examined for X-ray binaries in Fender \& Bright (2019).

Applying minimum energy analysis first to the optically thin peak, we may estimate a size scale of $c\Delta t$, although we note that this has large uncertainties. Using a flux density associated with the optically thin flare of 300 mJy at 5 GHz and a radius of the emitting region of $c\Delta t = 3 \times 10^{14}$ cm, we apply minimum energy analysis. The corresponding minimum energy is around $9 \times 10^{39}$ erg for an equipartition field of $\sim 0.04$G, the energy split equally between the relativistic electrons and the magnetic field (all assuming a filling factor of unity). Reducing the filling factor (the fraction of the inferred total volume responsible for the synchrotron emission, which is largely unknown), reduces in turn the energy requirement; for filling factor $f=0.1$ the energy reduces to $3 \times 10^{39}$ erg. 

For the peak optically thick flare(s) on day 11 the equipartition analysis (following Fender \& Bright 2019) under the assumption that the peaks were due to synchrotron self absorption gives us an estimated radius of $\sim 10^{13}$cm, internal energy $\sim 10^{39}$ erg and magnetic field of $\sim 2$ G, i.e. somewhat lower energy (but within a factor $<10$) and a much smaller size. If we assume that in the subsequent 24 hours the source continued to expand at $\sim 0.1$c (Tetarenko et al. 2017; Fender \& Bright 2019) then by the time of the optically thin peak one day later the radius of the ejecta responsible for the day 11 optically thick flare would be $\sim 3 \times 10^{14}$ cm, consistent with our estimate for the size of the optically thin shock region; i.e. the ejecta expanded physically by a factor of $\geq$ten between the optically thick peak and the later optically thin shock. Extrapolating the same expansion speed backwards implies the ejecta responsible for the day 11 peak were launched hours or less beforehand. We conclude therefore that a plausible scenario exists which is consistent with both source sizes and energetics: the ejecta responsible for the optically thick flare on day 11 interacted (likely decelerated) with local matter one day later, leading to the day 12 optically thin flare. These approximate arguments are not proof of this scenario, but they do demonstrate that it is plausible.

Finally we may also consider then, what happens if {\em all} the optically thick flares are decelerated, producing an optically thin event, approximately one day later. We can empirically see that the ratio of optically thick to thin peaks is $\sim 10$, and then inspect again Fig 2 which shows the large number of densely space optically thick flares. It seems likely that all of the subsequent optically thin events could be hidden by subsequent optically thick flares, but may contribute to the overall slow trend to more optically thin emission evident in the evolution of the spectral index between days 2 -- 10 zxx.

\subsection{Connection of radio emission to optical and X-ray bands}

We can compare our radio measurements somewhat more quantitatively to the contemporaneous X-ray and radio data already plotted in Figs 2, 6 and 7. In Fig {\ref{correlations} we plot the Pearson correlation coefficient between the radio (AMI), X-ray (ISGRI) and optical bands (i.e. three two-way correlations) as a function of binning (averaging) time (binning is required in order to identify quasi-simultaneous points).

\begin{figure}
\epsfig{file=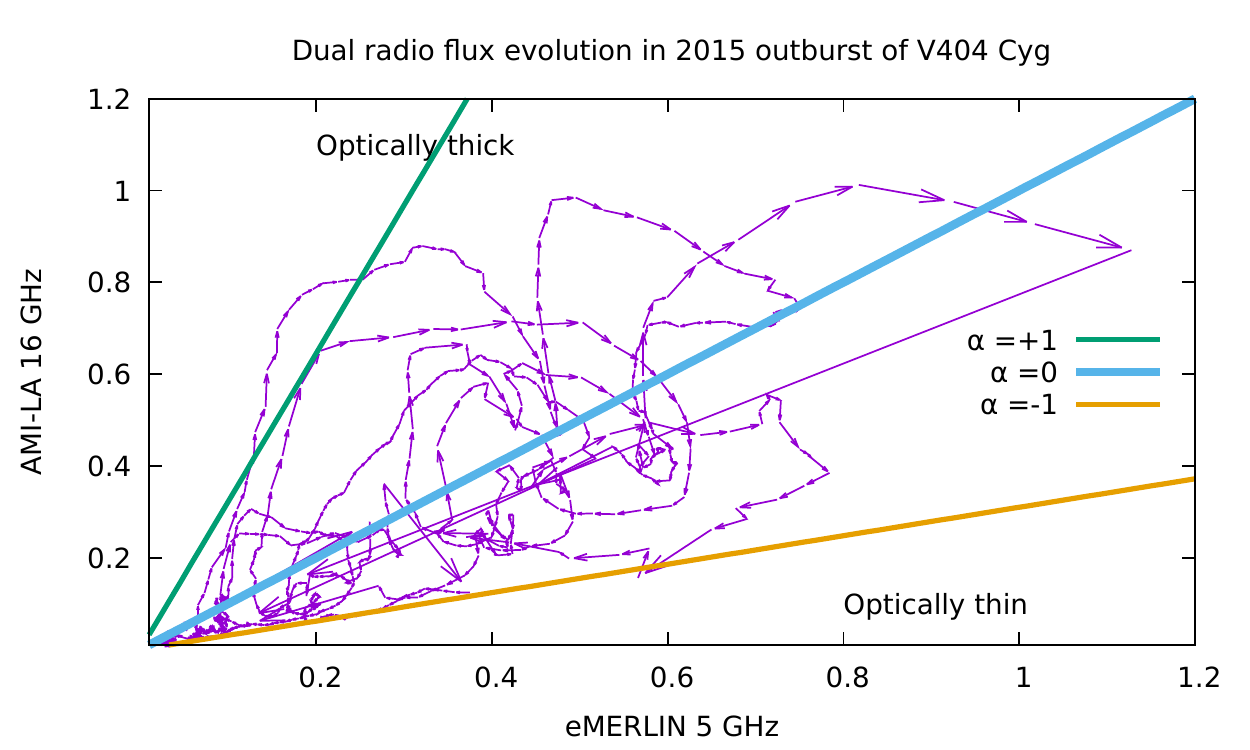, angle=0, width=8.5cm}\\
\epsfig{file=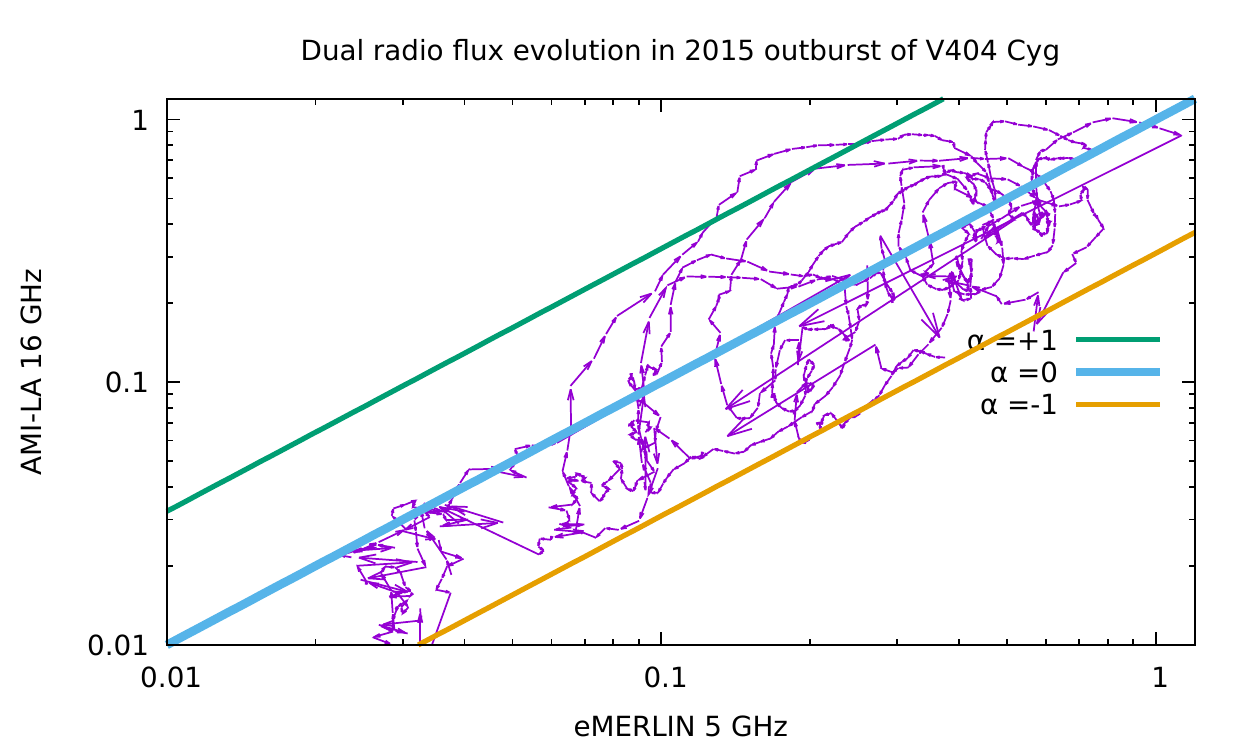, angle=0, width=8.5cm}
\caption{Vector plots of the radio-radio plane, illustrating the evolution of the radio flaring activity observed in the two radio bands (5 and 16 GHz), using data averaged to x-second bins. Diagonal sold lines illustrate instantaneous spectral indices of +1, 0 and -1 (green, blue and orange, respectively). Repeated clockwise loops are observed, corresponding to earlier and stronger peaks at the higher frequency, as the optical depth of the source evolves. Nearly all spectral index measurements fall in the range -1 to +1. Upper panel: linear-linear, lower panel: log-log.}
\label{loops}
\end{figure}

\begin{figure}
\epsfig{file=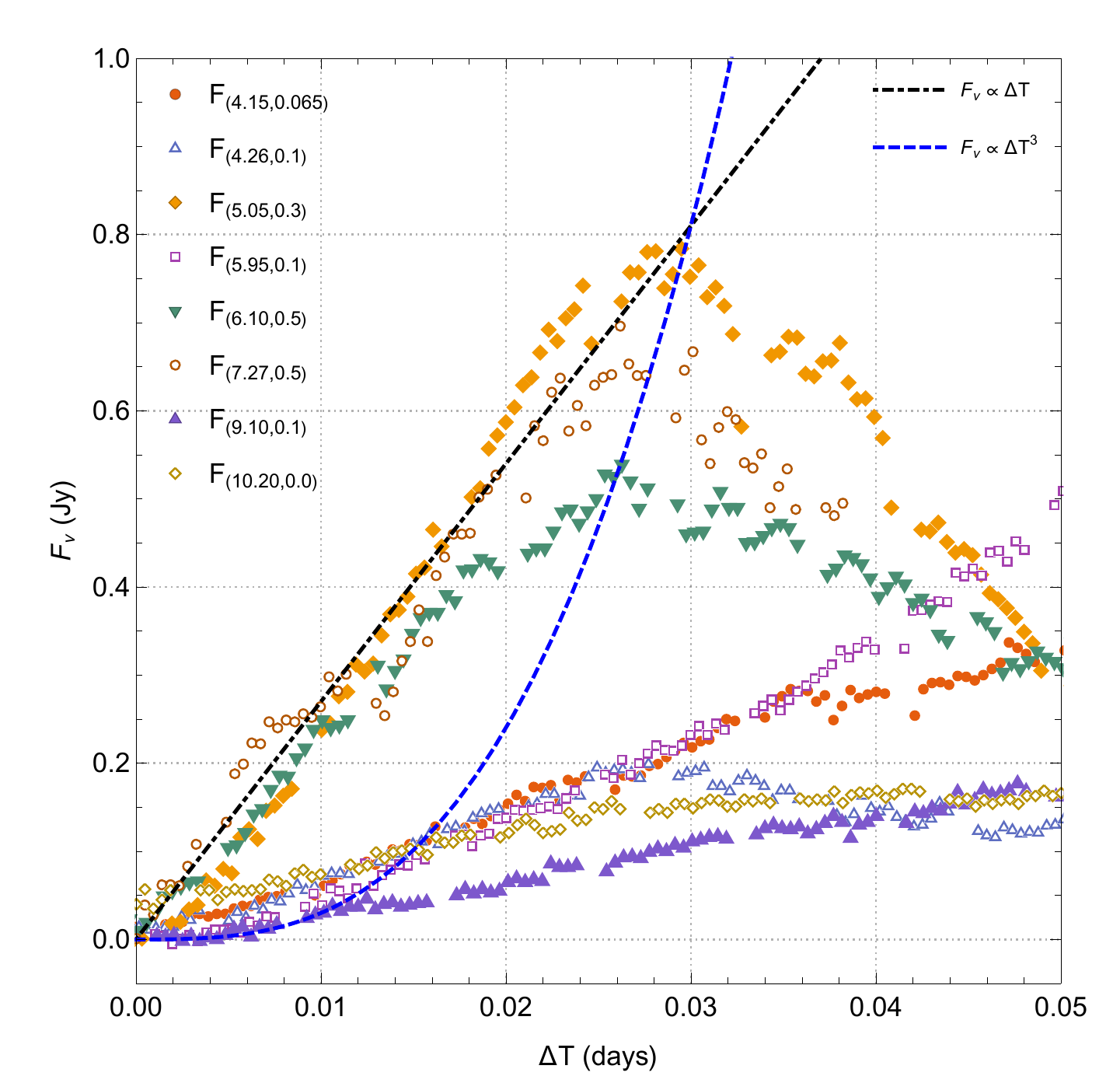, angle=0, width=7.5cm}\\
\epsfig{file=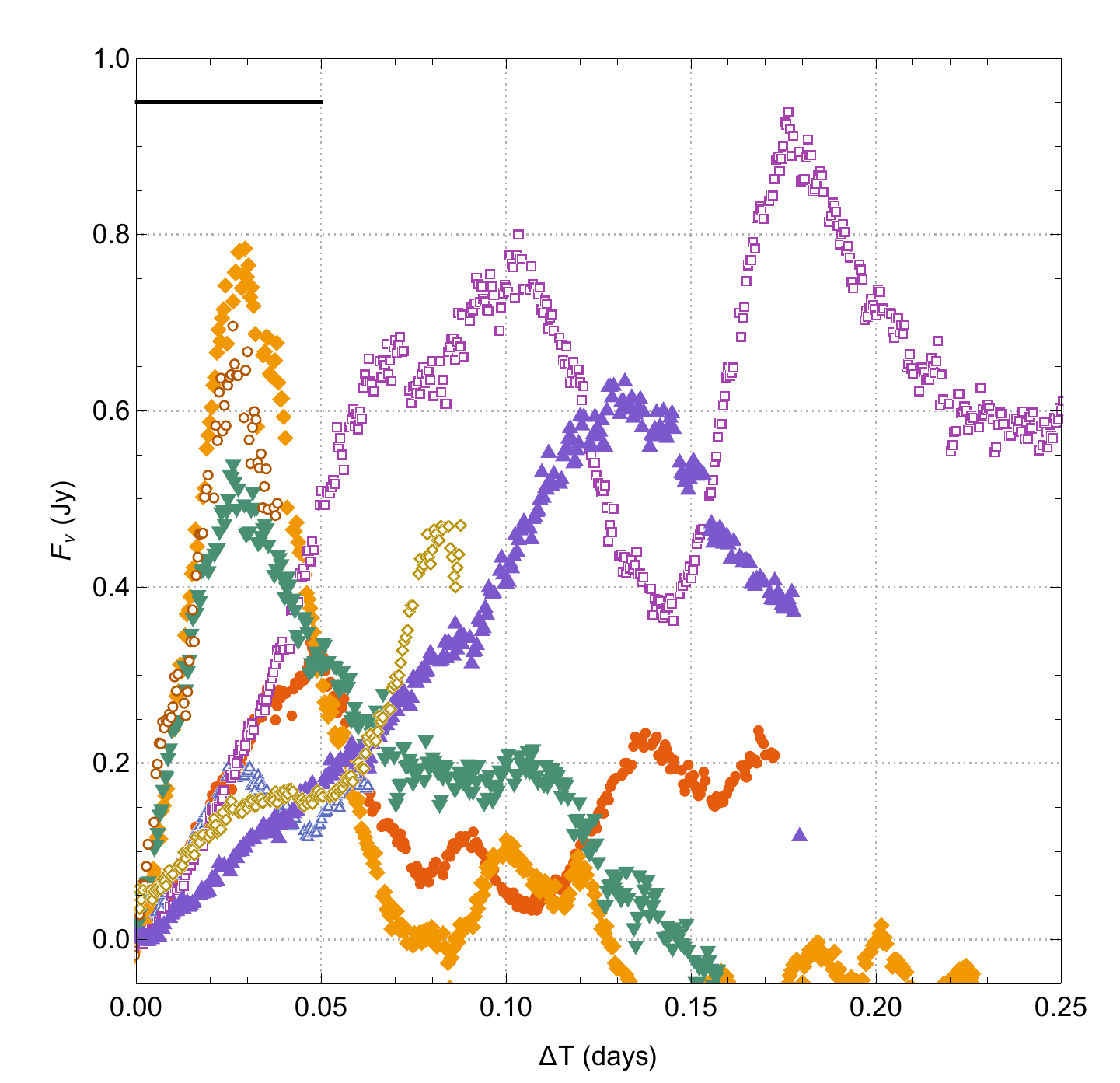, angle=0, width=7.5cm}
\caption{Upper panel: Eight specific radio flares, at 16 GHz, examined in detail. The two numbers associated with each flare in the legend indicate (i) the start time of the event (MJD 57188.0+first number) and (ii) subtracted baseline flux (Jy) for each flare. On timescales of 0.05 day there appears to be evidence for a dichotomy in the rise rates of the radio flares, with some events rising twice as fast as others (for a linear fit). Overplotted are a relation of the form $F_{\nu} \propto \Delta T^3$ as expected in a simple van der Laan (1966) expanding blob model, as well as a linear growth of flux density with time, which appears to fit the data better (neither line has been fit to the data).
Lower panel: the same events as in the upper panel, but shown over a longer timescales (the solid horizontal line indicates the time period covered in the upper panel). We see that the faster-rising group of events tended to be essentially isolated events, while the slower-rising events tend to be part of temporally-broader structures, most of which show abrupt changes in their rise rates at different times, and could not be fit by simple van der Laan-style models (see later). }
\label{rise1}
\end{figure}

\subsubsection{Connection between radio and X-ray emission}

In this paper we compare the radio data to X-ray data from INTEGRAL, specifically JEM-X1 (5 -- 10 keV) and ISGRI (25 -- 200 keV).
We find that the overall connection between X-ray and radio emission is considerably harder to discern than it has appeared in other X-ray binaries, probably due to the very large contribution from strong and variable absorption in the X-ray band (Motta et al. 2017a, 2017b).
The highly irregular nature of the outburst, comprising many large-amplitude flares, is reflected in both the radio and X-ray light curves, which differ in the same way (qualitatively) to `normal' outbursts. It has been previously observed that radio emission often responds to changes in X-rays with a typical time delay of tens of minutes in X-ray binaries (e.g. Mirabel et al. 1998; Klein-Wolt et al. 2002) and indeed this has already been reported explicitly for the decay phase of this outburst (Plotkin et al. 2017). This response probably represents a direct connection between accretion states and jet modes (on/off, steady/transient), with the delay corresponding to the particle acceleration timescale and/or the timescale for the synchrotron-emitting ejecta to become optically thin (during which time it is flowing downstream in the jet); see e.g. discussions in Mirabel et al. (1998); Klein-Wolt et al. (2002); Fender et al. (2004). Inspection of Fig \ref{ami-mer} shows several examples where the radio and X-ray appear to be strongly correlated on such timescales. However, for the V404 Cyg data presented here establishing a direct connection between individual X-ray and radio events is very difficult. This is due to a number of reasons, including: 
the incomplete overlap between the X-ray and radio emission, the large number of events in both bands, and the knowledge that the radio emission we observe may already be blends of multiple flares seen in the millimetre band, which are also smoothed and reduced in amplitude already at radio frequencies (Tetarenko et al. 2017).
Unfortunately we do not have good X-ray coverage around the times of the brightest radio events (Fig \ref{flare}) but the long, hard, X-ray dip, lasting about a quarter of a day, is notable (but bear in mind this may be absorption -- Motta et al. 2017a,b).

If we consider now Fig \ref{correlations}, we see that the radio and X-ray data are correlated on all binning timescales up to one day (above one day they are also almost certainly correlated, but the sampling pattern dominates the correlation). A power-law fit to the correlation results in a fit of the form $b = 0.43 \pm 0.06$ ($7.7 \sigma$ significance; where $F_{\rm AMI} \propto R_{\rm ISGRI}^b$ and $R_{\rm ISGRI}$ is the measured count rate) for most timescales. One example, with this fit, is presented in Fig \ref{xr}; we note the relatively large scatter around the fit. This fitted slope is consistent for that reported historically for the hard state in V404 Cyg (Corbel, Koerding \& Kaaret 2008) and also during the decline of this outburst (Plotkin et al. 2017). The implication here is that poor X-ray and radio sampling of a similar source, with lower flux, would have looked like a typical hard state radio:X-ray correlation, although the detailed observations show it is anything but.

\subsubsection{Connection to optical emission}

As with the X-ray data, we do not find convincing evidence for one-to-one correlation between individual optical and radio events. However, averaging the radio, X-ray and optical data over a range of timescales from 10 minutes to one day, as also presented in Fig \ref{correlations} we find the following:

\begin{enumerate}
\item Radio and X-ray are always correlated, with a relatively flat level of correlation on all timescales (as noted above). 
\item Radio and optical are not correlated on the shortest timescales, but have a rising correlation coefficient with averaging timescale, peaking at $\geq 500$ minutes.
\item X-ray and optical are similar to (ii) but reach the peak earlier, at $\geq 300$ minutes. This correlation also appears to weaken at the longest timescales (above $\sim$800 minutes).
\end{enumerate}

\begin{figure}
\centerline{\epsfig{file=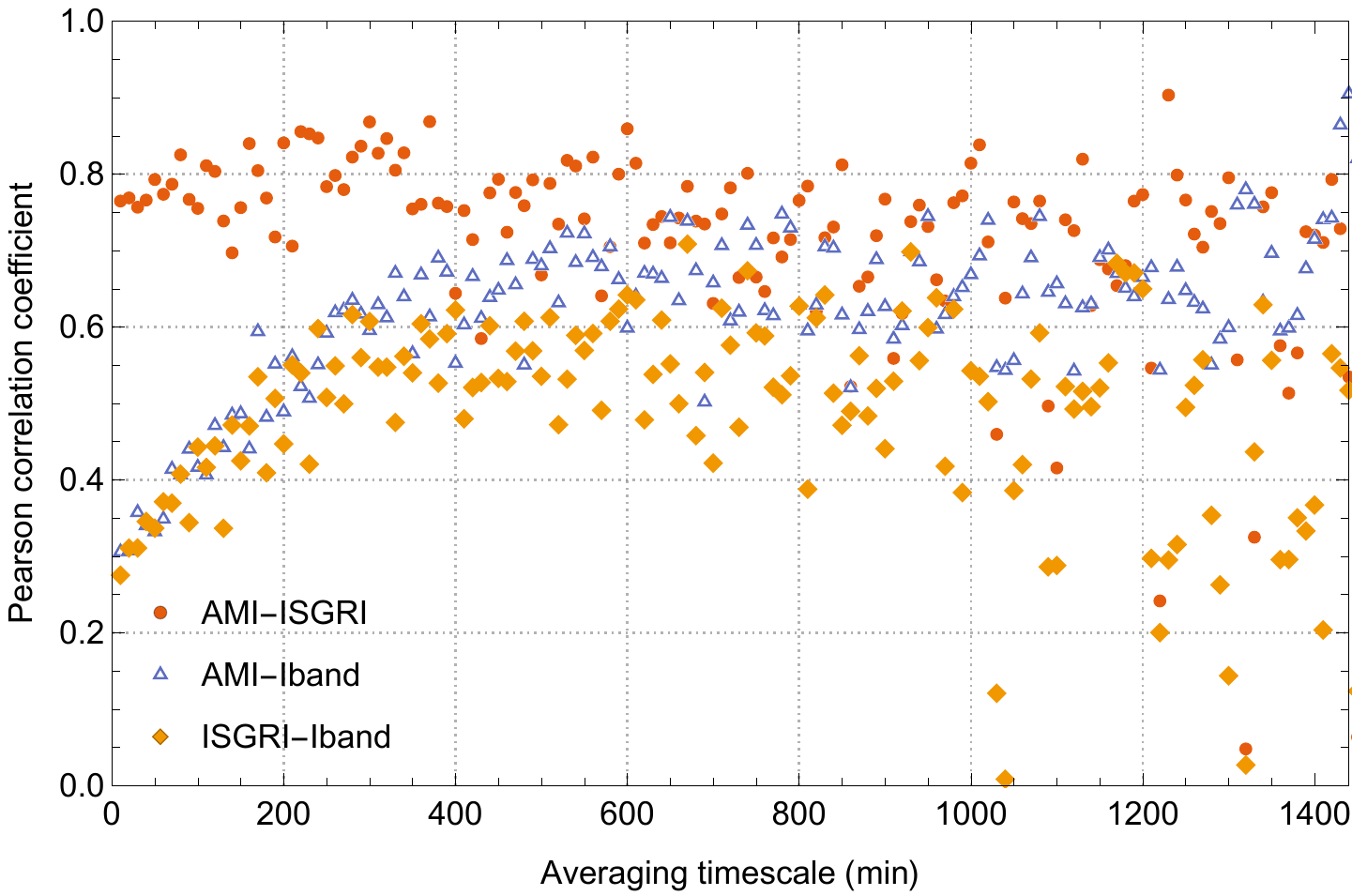, angle=0, width=8cm}}
\caption{Correlation between X-ray, optical and radio fluxes. The figure shows the Pearson correlation coefficients for the $Log_{10}$ of each of the relevant fluxes, as a function of timescale (minutes) over which the data were binned to get simultaneous measurements. Above about one day the correlation is dominated by the sampling, so we do not show it here. The radio and X-ray flux densities are correlated on all timescales, with a similar slope for the correlation (see below). Correlations with the I band optical flux density are instead not very strong at the shortest timescales but reach maximum strength at around 2--400 minutes.}
\label{correlations}
\end{figure}

\begin{figure}
\centerline{\epsfig{file=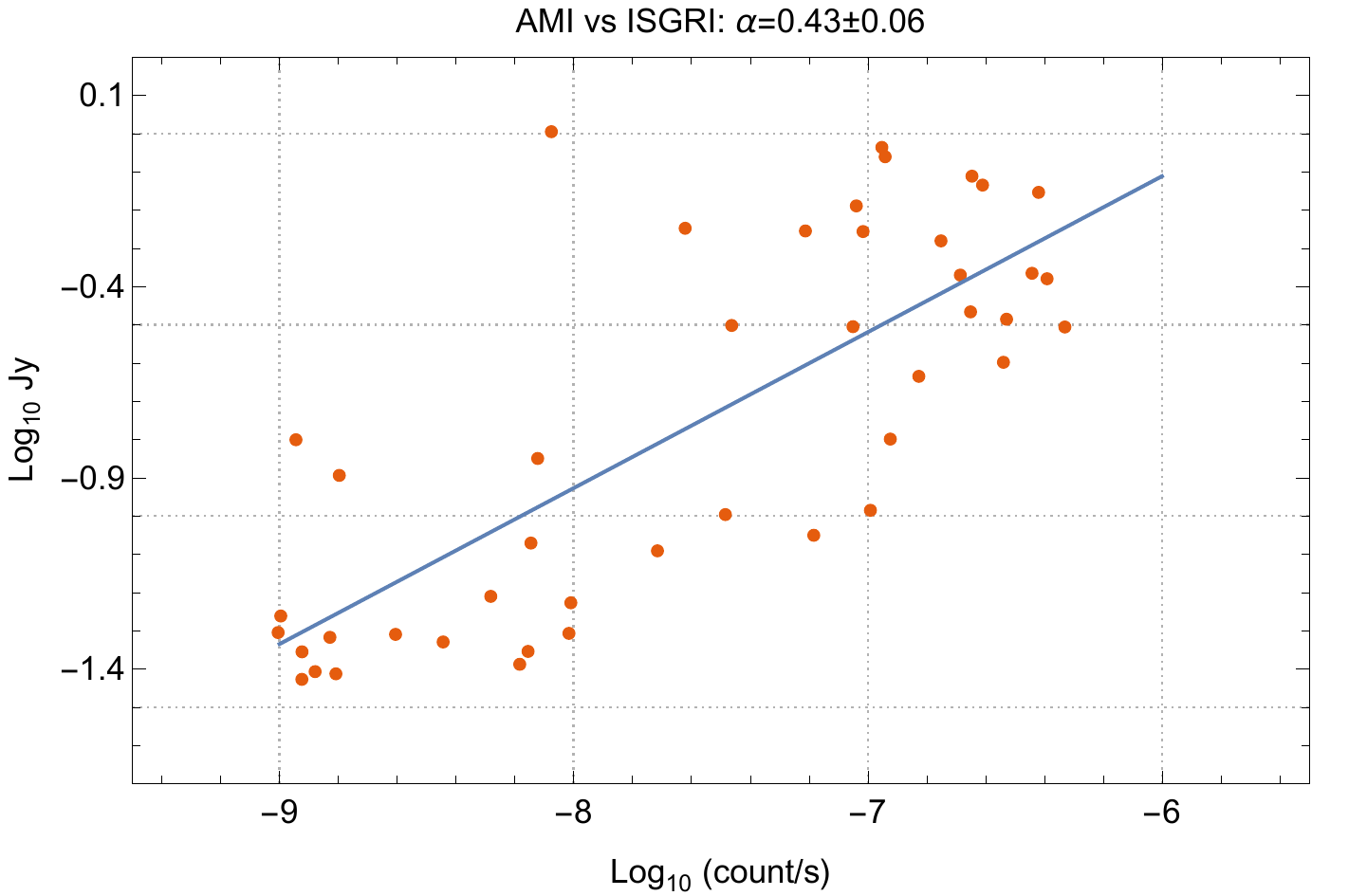, angle=0, width=8cm}}
\caption{Correlation between radio (AMI 16 GHz) and X-ray (ISGRI 25-200 keV) fluxes, made with 90-minute binning. The fitted slope of the correlation is always approximately of the form $F_R \propto F_X^{0.4}$ regardless of the binning.}
\label{xr}
\end{figure}

The slope of the moderate correlation between the radio and optical fluxes is similar to that of the radio and X-ray fluxes, with $b=0.47 \pm 0.15$ ($3.1 \sigma$) for $F_{\rm AMI} \propto F_{\rm I-band}^b$. We do not plot this correlation here, it is considerably weaker than that of the radio vs X-rays, itself weaker than in a number of comparable studies. The correlation between the optical and X-ray fluxes is similar, with 
$b=0.52 \pm 0.13$ ($3.9 \sigma$) for $F_{\rm I-band} \propto R_{\rm ISGRI}^b$. This value of $b$ is again very close to the global correlation between optical and X-ray fluxes in low-mass X-ray binaries reported in Russell et al. (2006).

Optical and near-infrared emission in X-ray binaries may arise from a combination of accretion disc viscous dissipation and irradiation (Charles \& Coe 2006), as well as from synchrotron emission near to the base of the jet (e.g. Fender \& Pooley 1998; this arises at distances at which the radio emission is self-absorbed). For this outburst, there is less correlation between the radio and optical data than between the radio and X-rays. What is very clear is, again, the very high levels of variability in optical as in the radio and X-ray lightcurves. The optical cycles reported in Kimura et al. (2016, whose data we use here) are clearly visible around day 5.7-5.9. Perhaps most strikingly, we see strong optical oscillations of increasing amplitude around day 10.7, the time when the strongest ejection events were taking place (Fig \ref{flare}). The very strong decline in the optical emission beginning around day 12 is also very obvious in Fig \ref{flare}.



\section{Kinetic feedback}

The varying radio synchrotron emission from X-ray binaries and other accreting sources traces the kinetic feedback from the system, and can be compared to other power/sink channels such as the radiation and advection (see e.g. Fender \& Munoz Darias 2016 for a recent review). There are two ways to approach a power estimate:

\begin{enumerate}
\item{Break the radio light curve down into a number of discrete synchrotron flares, and estimate the energy associated with each event}
\item{Assume a continuous power injection into thet jet, calculated directly from the core radio flux}
\end{enumerate}

The former method is more obviously applicable to V404 Cyg, which has a lightcurve dominated by flaring events. One major difficulty in extracting energy estimates from observations for such systems in general is uncertainty in the Doppler factor. In the case of V404 Cyg, however, the combination of a precise distance from radio parallax (Miller-Jones et al. 2009), a good inclination estimate from optical studies and measured proper motions on VLBI scales appears to constrain the bulk velocity to be at most mildly relativistic (Miller-Jones et al. 2019). We therefore assume a Doppler factor $\delta \sim 1$ and need apply no corrections to the estimates made below. This makes the situation considerably simpler than for nearly all other X-ray binaries, where the bulk velocity is essentially unconstrained (Fender 2003; Miller-Jones, Fender \& Nakar 2006).

\subsection{Feedback from flares}

We can take the distribution of flares automatically measured (Fig \ref{flaredist}) and use these to estimate the kinetic feedback associated with the flaring mode. In Fig \ref{peakshisto} we also show the estimated size, internal energy and magnetic field of the ejecta under the assumption that the peak was due to a transition from optically thick to optically thin to synchrotron self absorption. Specifically, we use equations (29) to (31) from Fender \& Bright (2019) (recasting their equation [29] to provide a size directly by multiplying by $c \Delta t$, where $\Delta t$ cancels and so does not have to be estimated). The distribution of peak flux densities is non-uniform and asymmetric enough that it is clear to see how the differing dependencies on it (stated in the box in each panel) variously compress, stretch and invert the distribution. The sum of the minimum energy associated with the 86 discrete flares observed is $3 \times 10^{40}$ erg, which we again may convert crudely to a real sum by dividing by 0.4 (our duty cycle of observations), leading us to an estimate of $7 \times 10^{40}$ erg.

\begin{figure}
\centerline{\epsfig{file=peakshisto.pdf, angle=0, width=7cm}}
\centerline{\epsfig{file=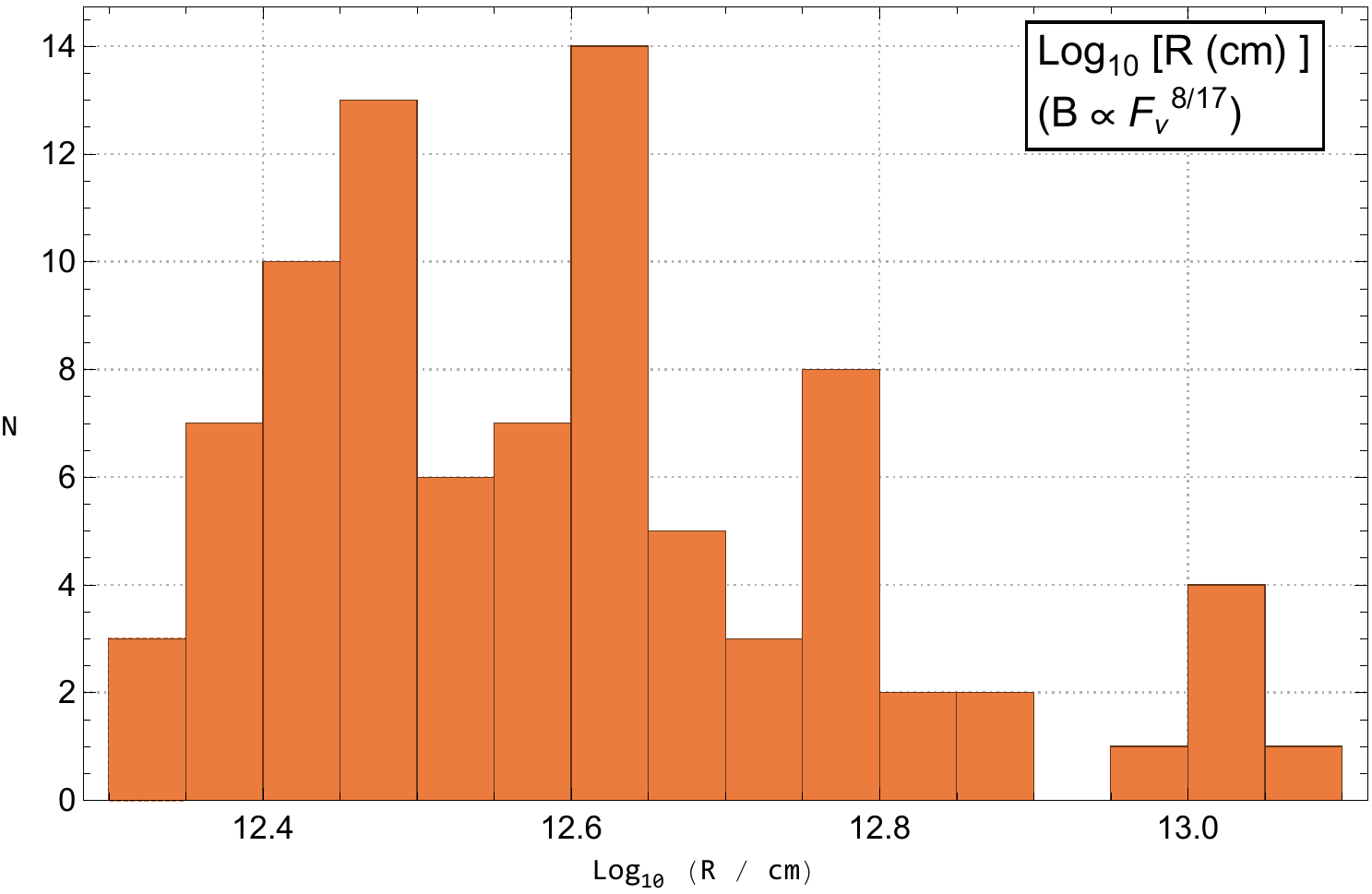, angle=0, width=7cm}}
\centerline{\epsfig{file=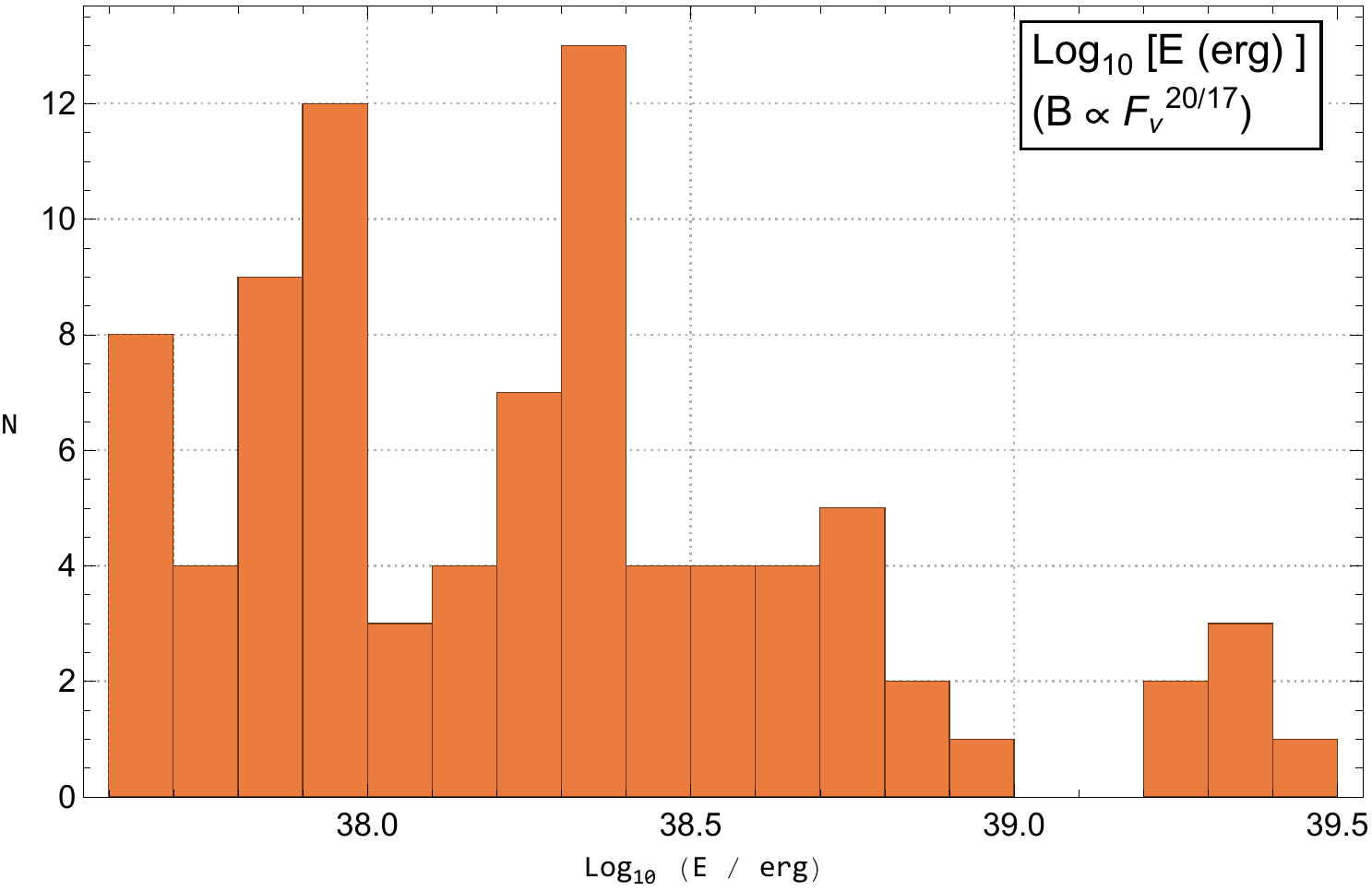, angle=0, width=7cm}}
\centerline{\epsfig{file=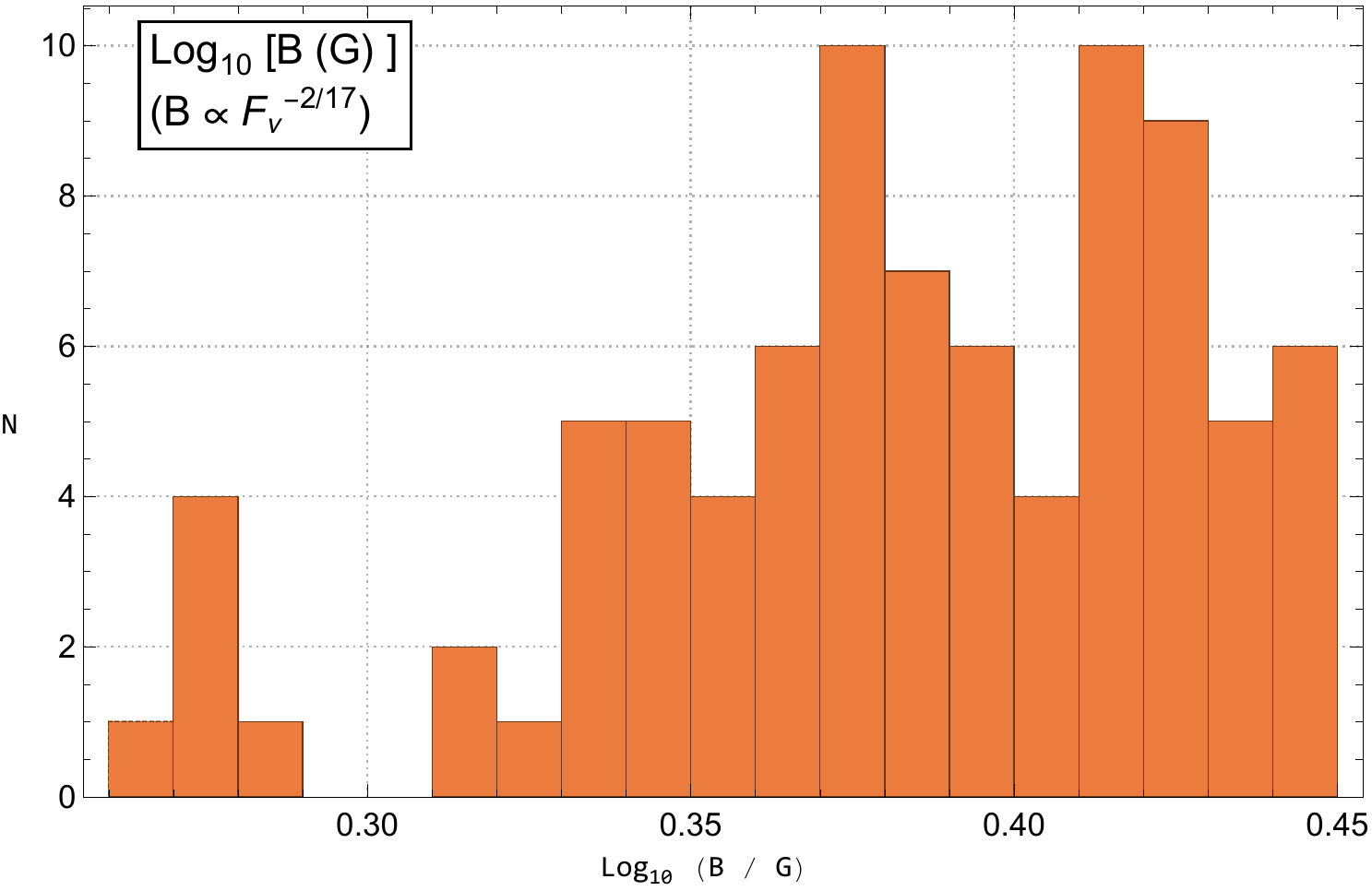, angle=0, width=7cm}}
\caption{Analysis of 86 flare peaks of $> 100$ mJy at 17 GHz. Top panel: flux densities at flare peak. Second, third and fourth panels indicate the physical quantities (size, minimum internal energy, magnetic field) inferred for the ejecta under the assumption of the peak arising due to the source becoming optically thin to synchrotron radiation at that frequency.}
\label{peakshisto}
\end{figure}

This sum is not, in the context of X-ray binaries, a particularly large number. Fender \& Bright (2019 and references therein) use exactly the same method to calculate the energy associated with radio flares from the X-ray binaries Cygnus X-3 and GRS 1915+105 (as well as two from V404 Cyg) and find that single large events from Cyg X-3 can have estimated internal energies $> 10^{41}$ erg. The size and magnetic field estimates are also consistent with other X-ray binaries. Note that studies such as Bright et al. (2020) have indicated that energy estimates from flaring are likely to be significant underestimates of the true total jet power. 

\section{Discussion}

We stated earlier in the paper that these observations represented the best ever coverage of radio flaring from a black hole X-ray binary in outburst. We can quantify this a bit more by making a comparison with GRS 1915+105, which has probably the best coverage prior to this data set (e.g. Klein-Wolt et al. 2002). In terms of probing to lower luminosities, the relevant observational metric will scale as $(S d^2)^{-1}$ where $S$ is the sensitivity (e.g. rms noise in one second) of the radio telescope, and $d$ is the distance to the source. A coverage parameter $C$, could variously be intensity of monitoring during a given period, if that was the goal of the comparison, or perhaps number of samples if long-term monitoring were the goal of the comparison. An approximate metric for comparing coverage is then $C (S d^2)^{-1}$.

\begin{figure}
\centerline{\epsfig{file=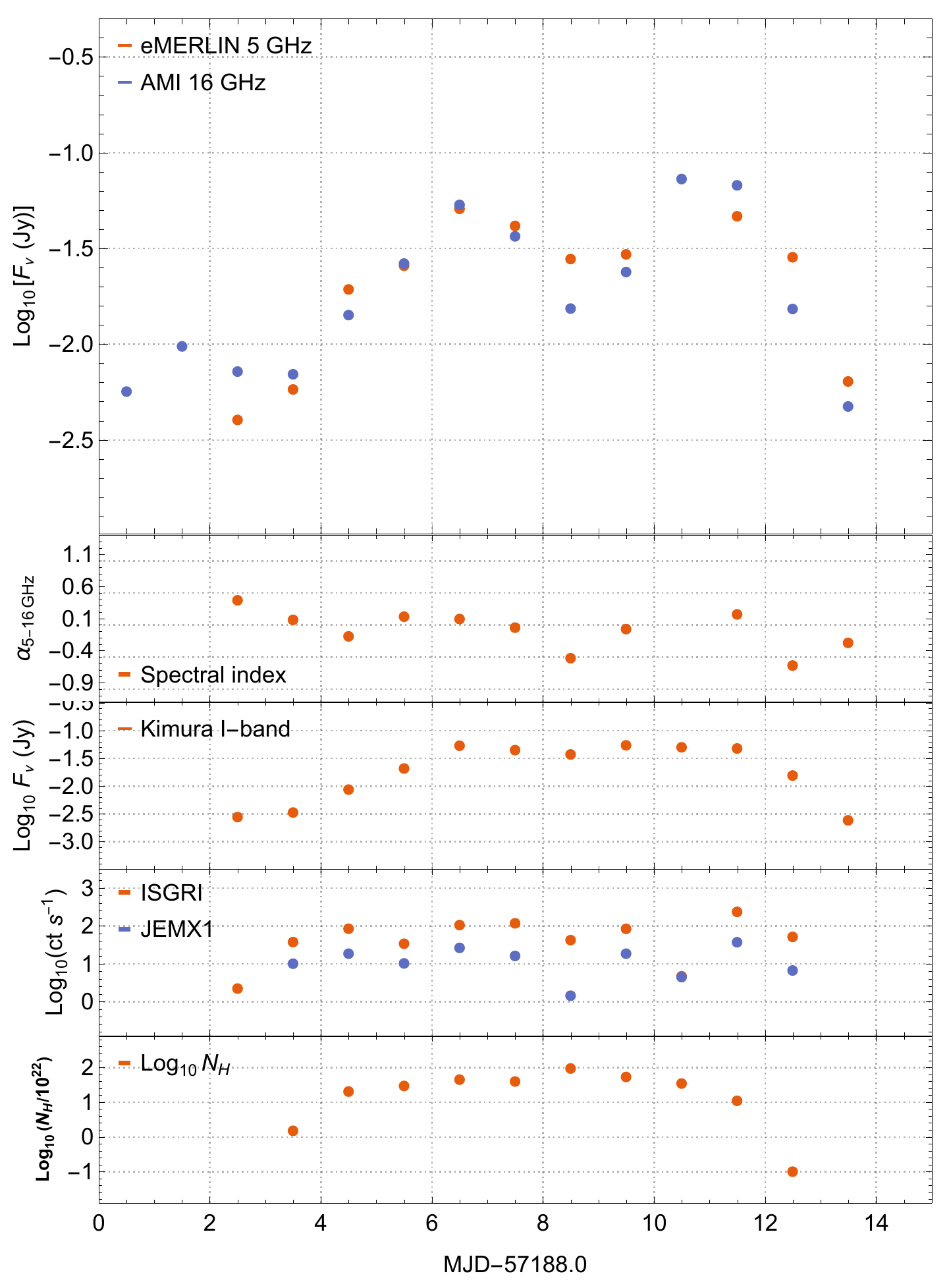, angle=0, width=8cm}}
\caption{The same as Fig 2, but with all the data averaged to one day timescales, typical of much of the sparse monitoring of black hole binaries which is available, particularly in the radio band. The flux densities have also been scaled to place the source at 8 kpc, which is a more typical distance for an X-ray binary.}
\label{1day}
\end{figure}

We can compare a typical observation of GRS 1915+105 with the Ryle Telescope (physically the same telescopes as AMI-LA, but with less bandwidth) with the AMI-LA observations of V404 Cyg presented here: $S_{RT} / S_{AMI-LA} \sim 3$. Considering the relative distances, $d_{1915} / d_{V404} \sim 4$. Finally, if we consider candence of coverage over the two weeks of the outburst, $C_{1915} / C_{V404} \sim 3^{-1}$. The ratio of metrics, for intense coverage over a two-week period, is then $\sim 150$. Of course such a crude analysis does not consider the additional physics which is accessed via complementary multiwavelength observations, nor the peculiarities of an individual source or outburst, but it does demonstrate that the intensive coverage of V404 Cyg in June 2015 probed the radio emission far more comprehensively than ever previously achieved.

It is instructive to consider what the data would have shown us if we only had daily (or worse) sampling, which is quite typical for X-ray and radio (especially) measurements, and had the source been at a more typical distance for an X-ray binary of 8 kpc. Fig \ref{1day} reproduces Fig 2 but with the sampling averaged to one day and the fluxes reduced by a factor $(2.4/8)^2$. As the figure is only meant to be illustrative, there is little point in estimating error bars - the measurements can be readily compared with observers' own knowledge of their instruments; furthermore the $N_H$ and radio spectral index are not affected. What we see is a radio source varying relatively smoothly but revealing three peaks, associated with a spectral index which varies in the usual range for a source with contributions from both optically thick and thin synchrotron emission. 
Inevitably all of the short timescale variability evident in every band, as well as in the inferred $N_H$, is lost. Less inevitably, broad correlations are evident between the radio and optical fluxes and the $N_H$. The former are not so surprising, the latter is perhaps more noteworthy, but we suspect does not have a deep physical significance.

Similar data sets (albeit usually without the varying $N_H$) have been, and continue to be, analysed as new sources are discovered and known sources (like V404 Cyg) go into outburst and are observed with far poorer sampling than we managed to achieve in this case. The multiple radio peaks would have been interpreted as evidence for multiple ejection events, supported by the variable radio spectral index. If we had only these data to analyse, and made the assumption that the three broad peaks of amplitude 50-100 mJy we saw were due to synchrotron self absorption, we would have concluded that the kinetic feedback was around $10^{39}$ erg, or about $\sim 5$\% of the figure we estimate from our higher-cadence data (using the same approach as previously, for a distance of 8 kpc). This clearly demonstrates the importance of sampling (and observing frequency) in estimating numbers and strength of radio flares and hence kinetic feedback (see also a similar discussion in Tetarenko et al. 2017).

\begin{figure}
\centerline{\epsfig{file=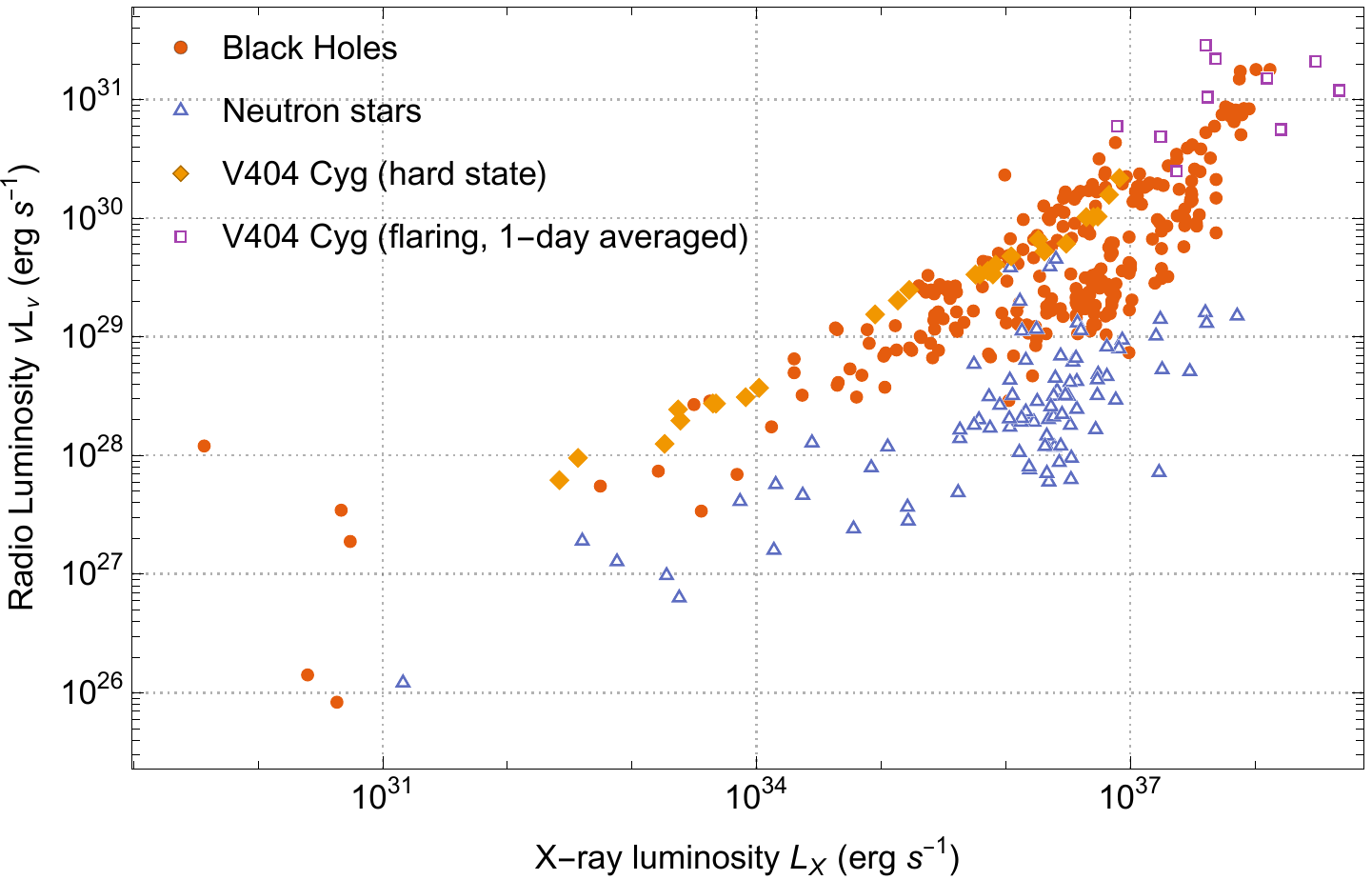, angle=0, width=8.5cm}}
\caption{The radio:X-ray plane comparing the one-day averaged points to the ensemble of hard state BH and NS from Bahramian (https://github.com/bersavosh/XRB-LrLx\_pub). The averaged flaring data scatter around the top end of the radio loud hard state correlation and are consistent, with some noise, with the published radio:X-ray correlation for V404 Cyg.}
\label{lrlx}
\end{figure}

\subsection{The need for an extended phase of particle acceleration}

In section 3.2.1 and Figure 5 we highlighted the failure of the simple van der Laan model to account for the ratios of flare peaks at different frequencies. Specifically, the lower frequency peak is much stronger than would be expected based upon the preceding high frequency peak. This issue was already recognised nearly 50 years ago in the case of the 1972 radio outbursts of the exotic X-ray binary Cygnus X-3. Peterson (1973) identified the same issue with the ratios of the peak flux densities, and demonstrated a simple solution: namely that there is an extended phase of particle acceleration, and there are more synchrotron-emitting electrons at the time of the second peak compared to the first. This is in contrast to the van der Laan model in which all of the energy and particles are injected impulsively. From this point much of the focus in the physics of radio variability from supernovae and AGN turned towards blast waves and the radio emission associated with propagating shock fronts (Blandford \& McKee 1977; Chevalier et al. 1982; Sari, Piran \& Narayan 1998). These models are very different from van der Laan primarily in that they invoke ongoing {\em in situ} particle acceleration and magnetic field amplification via the conversion of the kinetic energy of the blastwave. Such deceleration and interaction with the ambient medium is indeed likely to happen in the case of X-ray binary jets, and has been directly observed on large scales (e.g. Corbel et al. 2002; Bright et al. 2020; Carotenuto et al. 2021). Both types of models for extended phases of particle acceleration (whether related to the injection/launch phase of the jet, or the deceleration of the resulting shock wave) are able to produce the much flatter peak ratios we observed in V404 Cyg and other X-ray binaries. Note that Shahbaz et al. (2016) already report direct evidence for shocks in the jet of V404 Cyg during this outburst (although even the van der Laan model requires at least one initial shock or other particle acceleration mechanism).

We conclude that it is time that studies of jets from X-ray binaries adopted such blast wave models. Indeed it is ironic that the field of gamma-ray bursts has very developed theory for such events despite a paucity of high quality data, and in the field of jets from stellar-mass black holes and neutron stars we find ourselves in the inverse position: superb data being tackled with old models. We state this despite the apparent success of a modified van der Laan model being applied to a mm-cm flaring period in this source by Tetarenko et al. (2017).

However, beyond simply learning the lessons from parallel fields, X-ray binaries uniquely offer another possibility, which is that there is an extended, and measurable, period of injection of relativistic electrons from the inner accretion disc. In GRBs, SNe and related sources to which blast wave models are applied the energy injection phase is assumed impulsive, and in any case is much shorter than the shock acceleration phase. In X-ray binaries this is not always the case: we know (e.g. GRS 1915+105 in Klein-Wolt et al. 2002) that the timescales of X-ray states and state changes are comparable to the rise and decay times of the rapid flaring seen in some sources. Given the exceptional X-ray timing observations of X-ray binaries, we may well be able to directly tie the duration of an inferred particle injection event to observable behaviour of the accretion flow, convolved with the response of the resulting shock wave (see e.g. discussion in Wood et al. 2021).

\subsection{A comparison with GRS 1915+105}

In searching for an explanation for the anomalous nature of the June 2015 outburst of V404 Cyg, an obvious comparison to draw is with the archetypal source in the study of powerful jets and their connection to accretion states, GRS 1915+105. These two systems represent, to the best of our knowledge, the largest two black hole X-ray binaries in our galaxy (in terms of orbital separation, and hence accretion disc size), see table in e.g. McClintock \& Remillard (2006). There are certainly strong similarities: both sources show strong, quasi-oscillatory, behaviour in their X-ray emission (although it is more regular in GRS 1915+105, see e.g. Belloni et al. 2000), and present radio light curves which are comprised of very large numbers of flares. In both sources, active periods are ended by large radio flares (e.g. Fender et al. 1999). However, GRS 1915+105 did not until $\sim$ 2019 show the strongly-variable local absorption associated with V404 Cyg (Munoz-Darias et al. 2016; Motta et al. 2017a,b, 2021), and shows much more regular cycles of X-ray and coupled radio emission (e.g. Klein-Wolt et al. 2002). 

There are also, however, some dramatic differences. The first, relates to the duration of the outburst: GRS 1915+105 (33-day orbit) has been in outburst for nearly a quarter of a century (it was discovered in 1992: Castro-Tirado, Brandt \& Lund 1992), V404 Cyg (6.5-day orbit) was in outburst for only a few weeks. Neither source shows `typical' outburst behaviour, which can in fact be seen (more or less) in the third-longest orbital period system GRO J1655-40 (2.6 day orbit), whose 2005 outburst lasted a few months and followed an anticlockwise track in the HID (e.g. Fender, Homan \& Belloni 2009). Of course the very strong and variable absorption make the underlying accretion state of V404 in June 2015 hard to track (Motta et al, 2017a,b; but see e.g. discussion in Walton et al. 2017).


\subsubsection{The extended missing jets}

V404 Cyg does not show any resolved radio emission on scales greater than a few tens of milliarsec, i.e. those accessed by VLBI observations (Miller-Jones et al. 2019). The eMERLIN observations presented in this paper are the next highest angular resolution after those VLBI observations, and yet at no time do we see clear evidence for resolved ejecta. An interesting comparison between the two sources, with an emphasis on this aspect, can be made by considering the outburst and resolved ejecta from GRS 1915+105 reported in Fender et al. (1999), in which the same telescopes (both pre-upgrade) were used as in this study. In this case a radio flare from GRS 1915+105, peaking at $\sim 600$ mJy at 15 GHz follows approximately 20 days of a quasi-steady radio-bright hard state, and is itself followed by (at least) four days of strong radio oscillations. Multiple sets of radio ejecta were resolved by MERLIN out to 300 milliarsec, and previous observations with the VLA (e.g. Mirabel \& Rodriguez 1994; Miller-Jones et al. 2007) have resolved ejecta to $\sim$ one arsecond (if they did not become resolved out). In very few (if any) cases are the ejecta themselves spatially resolved (in particular perpendicular to the flow axis). GRS 1915+105 is about four times as distant as V404 Cyg, which means that the same ejecta could potentially have been observed to $\sim 3$ arcsecond, and equally that the VLBI observations of V404 Cyg (Miller-Jones et al. 2019, Wood et al. in prep) are probing a physical regime in which ejecta have not been previously resolved.

Why, then, do we not see the ejecta in V404 Cyg with eMERLIN? Two possible solutions present themselves: (a) the ejecta do not physically make it very far from the core, (b) the ejecta physically propagate far enough but are too faint to be observed, either because they are unresolved but have faded below the point-source sensitivity of the images, or because they are resolved and do not have sufficient surface brightness.

Is it plausible that the ejecta do not make it to the 40 milliarcsec scale which corresponds to the eMERLIN angular resolution? At 2.4 kpc, this corresponds to a scale of $10^{15}$cm. A somewhat quantitative comparison can be made taking into account the following:

\begin{itemize}
\item{eMERLIN, with an angular resolution of 40 mas, observed V404 Cyg between 2 and 14 days after the first radio flaring. On every day in between there was radio flaring, at varying levels (and strongest during the first ten days). Therefore, if the ejecta are propagating freely away from the binary, the eMERLIN observations correspond to times when ejecta are between just launched and 12 days old}
\item{The jets observed with VLBI (Miller-Jones et al. 2019) are observed to have proper motions of between 4 and 46 mas/day}
\item{Combining these facts, the eMERLIN images could have contained jets on angular scales between zero and 552 mas from the binary. Even the slowest-moving of the jets from the first epochs should have been resolvable by the last eMERLIN epochs}
\item{During days 3 -- 9, at least, there are radio flares typically occurring on intervals of an hour or less (and these themselves may be composed of a larger number of events on temporally-resolvable at higher frequencies, see Tetarenko et al. (2017) and Miller-Jones et al. (2019). The angular separation between these events, based on the observed VLBI proper motions, would be less than 2 mas}
\end{itemize}

Considering the above, the ejecta should have been at resolvable distances from V404 Cyg at at most ten days, and more likely within a few days, of the start of the outburst. Could the ejecta have been resolved out by eMERLIN? Our minimum-energy calculations in Fender \& Bright (2019) and above in this paper, lead us to estimate that the ejecta are expanding at around $0.1c$ and have reached physical sizes of $10^{12}$-$10^{13}$ cm after $\sim 0.2$ days. Continuing to expand at $0.1c$, the ejecta would reach radii comparable to the angular resolution of eMERLIN at around 4 days from launch, corresponding to extrapolated angular speeds of separation from the core of 8-92 mas/day. It seems likely, therefore, that we should have seen something with eMERLIN. Note furthermore that in the recent case of the jets from the black hole MAXI J1820+070, eMERLIN measured significant flux (whilst still resolving out 80\%) from ejecta months after launch (Bright et al. 2020), suggesting that some high surface brightness components persist.

\subsubsection{Choked jets and a nova-like shell?}

Based on the above discussion, we cannot rule out that the jets from V404 Cyg in 2015 did not make it very far from the launch site. This in turn suggests a link between these `choked jets' and the strong and variable local absorption of the source. Intriguingly, in the past two years GRS 1915+105 has entered a new, X-ray obscured state which shows further similarities to the obscured nature of the V404 Cyg outburst (Miller et al. 2020; Motta et al. 2021a). If the jets from V404 in June 2015 were not observed on scales larger than VLBI because they were being physically stopped beforehand, then they will have deposited their kinetic energy unusually close to the launch site (for comparison, we are talking about scales of $<10^{14}$ cm for V404 Cyg whereas many X-ray binary jets have been seen to propagate for distances of over $10^{17}$cm). Our estimate of the total kinetic feedback is $\sim 10^{41}$ erg, but (i) we have also shown that time-averaging can reduce the estimated kinetic power, and (ii) Tetarenko et al. (2017) have shown that at GHz frequencies we almost certainly {\em are} averaging over events. Therefore it is possible that considerably more than $10^{41}$ erg of kinetic energy was deposited in the environment close to the source. This overlaps with the estimated kinetic energy input into classical nova shells, which are known to persist for a long time, with clearly resolvable structures. We therefore suggest that a nova-like shell around V404 Cyg may have been created by the local deposition of so much kinetic energy. This kinetic energy input may also be associated with the nebular phase reported by Munoz-Darias et al. (2016).

\section{Conclusions}

We have presented extensive radio observations of the nearby black hole V404 Cyg during 15 days of bright outbursting activity in 2015. The data reveal a very large number of radio flaring events, similar to the behaviour observed in parallel in the X-ray and optical bands. For all but one of these flaring events, the spectral evolution is consistent with a peak due to synchrotron self-absorption, allowing us to estimate the internal properties of the ejecta, while the ratio of peak flux densities at different frequencies requires there to have been an extended period of particle acceleration rather than an impulsive injection. While this clouds the picture in terms of modelling, it does open up the possibility that in future we may be able to correlate finite-duration states in the accretion flow with such extended periods of particle acceleration. However, the possibility that the ejections {\em were} impulsive and the extended phase of particle acceleration is due to shocks in the flow cannot be ruled out. If this is the case then we should look to models for synchrotron emission from supernovae and gamma-ray bursts for inspiration in advancing our modelling.
Overall the data captured a snapshot of 100s, if not 1000s, of moderately powerful ejections (in the context of stellar mass black holes) of relativistic plasma from the black hole over this period. The lack of resolved jets with eMERLIN opens up the possibility that the mildly-relativistic jets observed on VLBI scales (Miller-Jones et al. 2019) did not make it beyond $10^{15}$ cm from the black hole, and we note that the local deposition of the integrated kinetic energy from the ejections could have produced something akin to a classical nova shell around the binary.

We investigated the properties of the distribution of fluxes and find a significant part of the distribution to be log-normal in form. This in turn led us to the discovery of the first reported flux-rms relation for radio emission from a black hole, with a near-linear correlation as also found in the X-rays.  This potentially opens up new avenues in understanding how signals propagate from the accretion flow to the jet. Averaging on a range of timescales from seconds to days, we find that the radio and X-ray fluxes are always significantly correlated, but that correlations with the optical flux are only strong on timescales of hours or above. 

The data also imply some lessons which can be learned. With such large amounts of data, and so many individual events, it becomes clear that determining, for example, one-to-one connections between X-ray and radio events is very difficult. Spurious connections and correlations are easy to find. In this case we were not able to confidently determine a single one, and conclude that it is much better to focus efforts on 'cleaner' sources such as GRS 1915+105 or other X-ray binaries. Averaging and correlating the data on a range of timescales clearly demonstrates that radio:X-ray correlations from flaring events can look very much like those we attribute to a `steady' hard state jet: the radio:X-ray and optical:X-ray correlation slopes for averaged flaring data are very similar to those for the hard state, as is the normalisation of the radio:X-ray relation (i.e. position in the radio:X-ray luminosity plane).  Furthermore, we will in general in observing such sources severely underestimate the number of flares (as already demonstrated clearly in Tetarenko et al. 2017) and hence will underestimate the jet power (there are exceptions to this rule: for some of the well spaced state transitions in GRS 1915+105 we can actually be quite confident we have counted every flare e.g. Klein-Wolt et al. 2002).

\section*{Data Availability}

The data underlying this article will be shared on reasonable request to the corresponding author.

\section*{Acknowledgements}

RF would like to acknowledge valuable conversations with Jorge Casares about the optical properties of V404 Cyg, with Lauren Rhodes and Alexander van der Horst about models for GRB radio emission, and with Adam Ingram and Phil Uttley about the radio flux-rms relation.  RF was partly funded by ERC Advanced Investigator Grant 267607 `4 PI SKY'. This research made use of many scientific software packages, including {\bf gnuplot}, {\bf scipy} and {\bf mathematica}. All of the radio data presented in this study are available for you to use for your own studies (please contact the authors).  eMERLIN, funded by the STFC, is a National Facility operated by the University of Manchester at Jodrell Bank Observatory. We thank the Mullard Radio Astronomy Observatory staff for scheduling and carrying out the AMI-LA observations. The AMI telescope was supported by the ERC under grant 307215 `LODESTONE', STFC, and the University of Cambridge. We thank the anonymous referee for useful feedback and comments.

\newpage
\newpage

\appendix
\section{Finding peaks}

We used the find\_peaks function from the publicly-available module \textbf{scipy signal} to search the AMI 17 GHz data for local peaks, with the following settings: (height = 0.1, prominence = 0.01, width=5, distance=10). Fig \ref{apppeaks} shows the fitted peaks from this method for the six intervals investigated in detail in the main paper and Fig 3. The fitted peaks correspond closely to those we identify by-eye.

\begin{figure*}
\epsfig{file=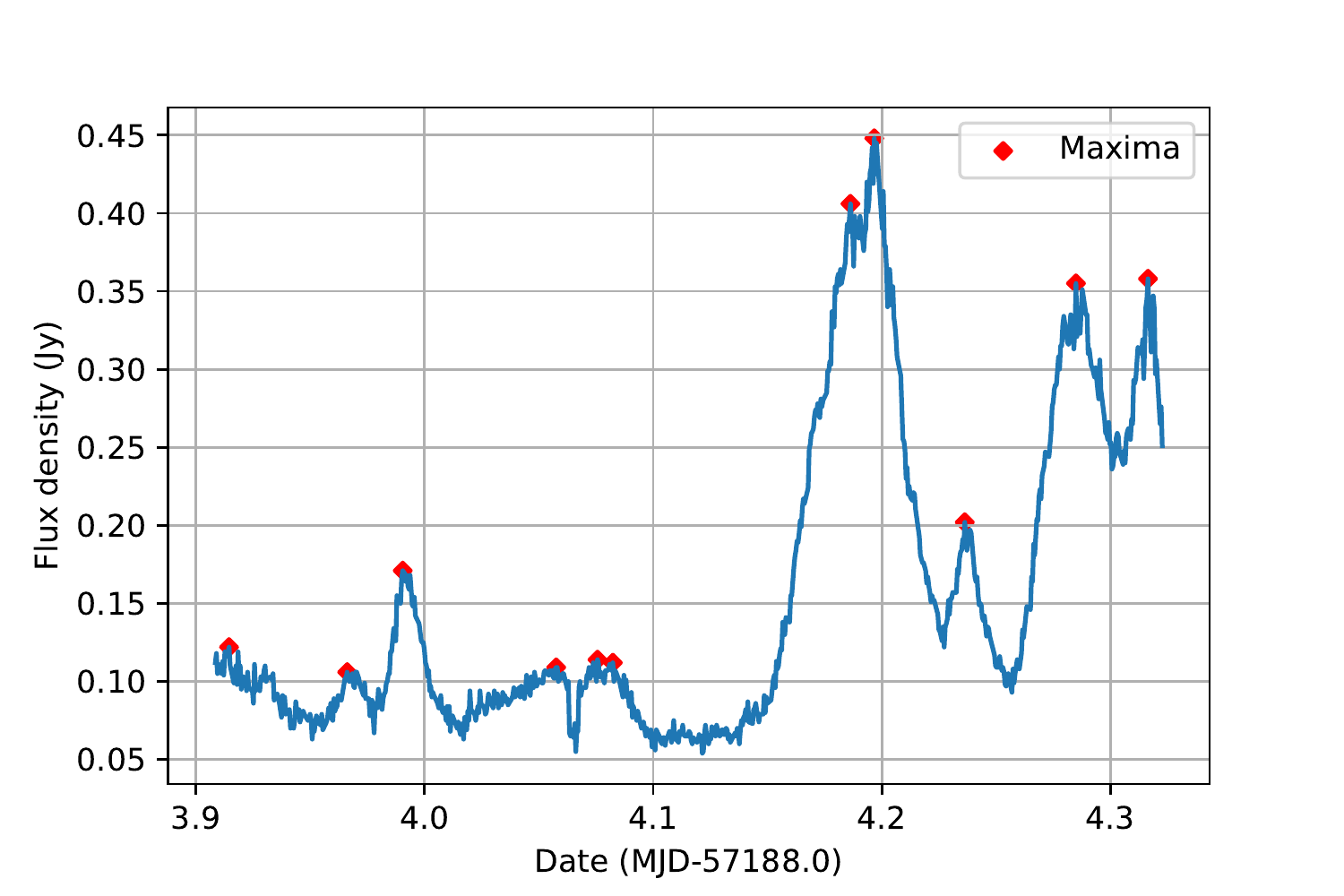, angle=0, width=7cm}\quad\epsfig{file=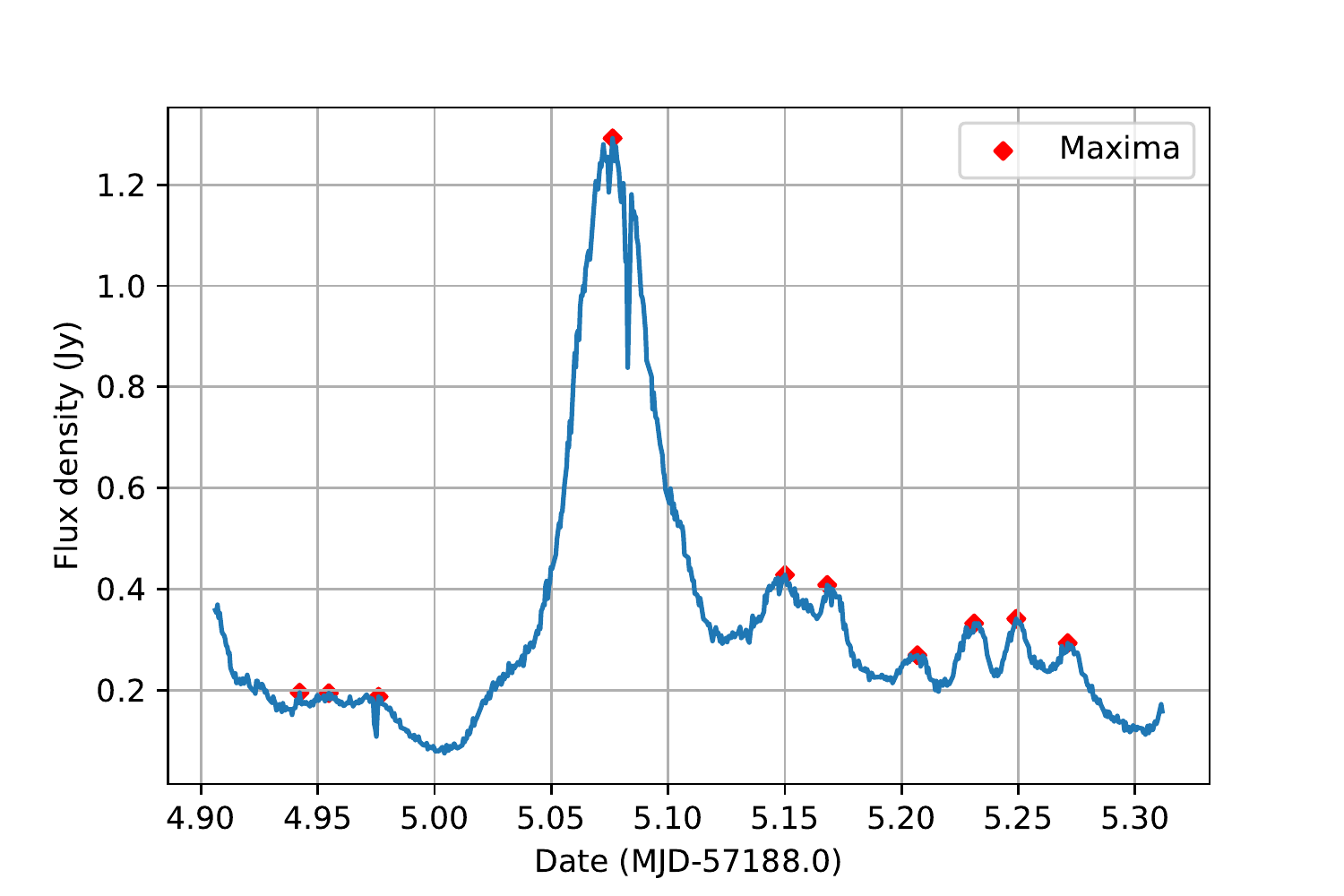, angle=0, width=7cm}\
\epsfig{file=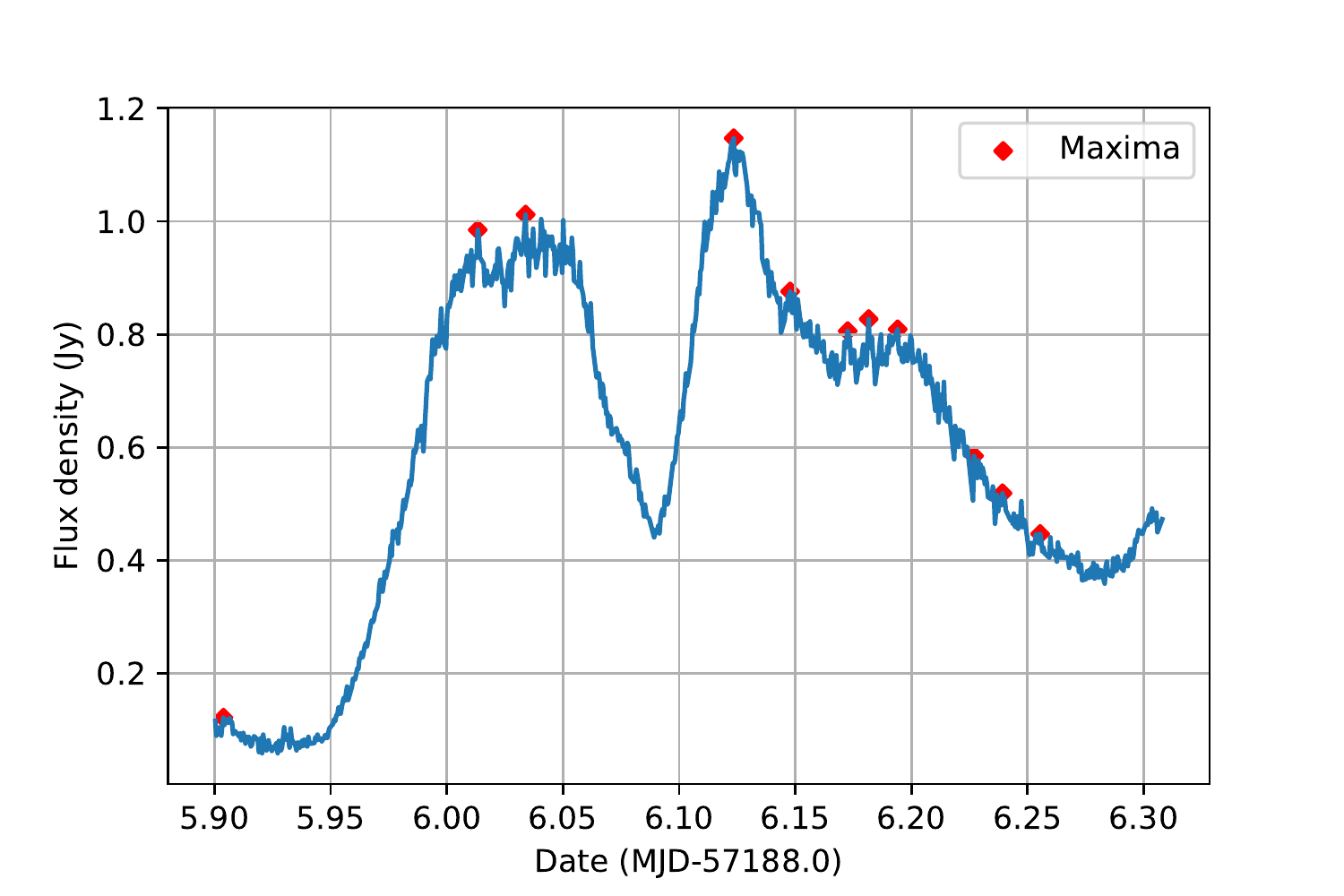, angle=0, width=7cm}\quad\epsfig{file=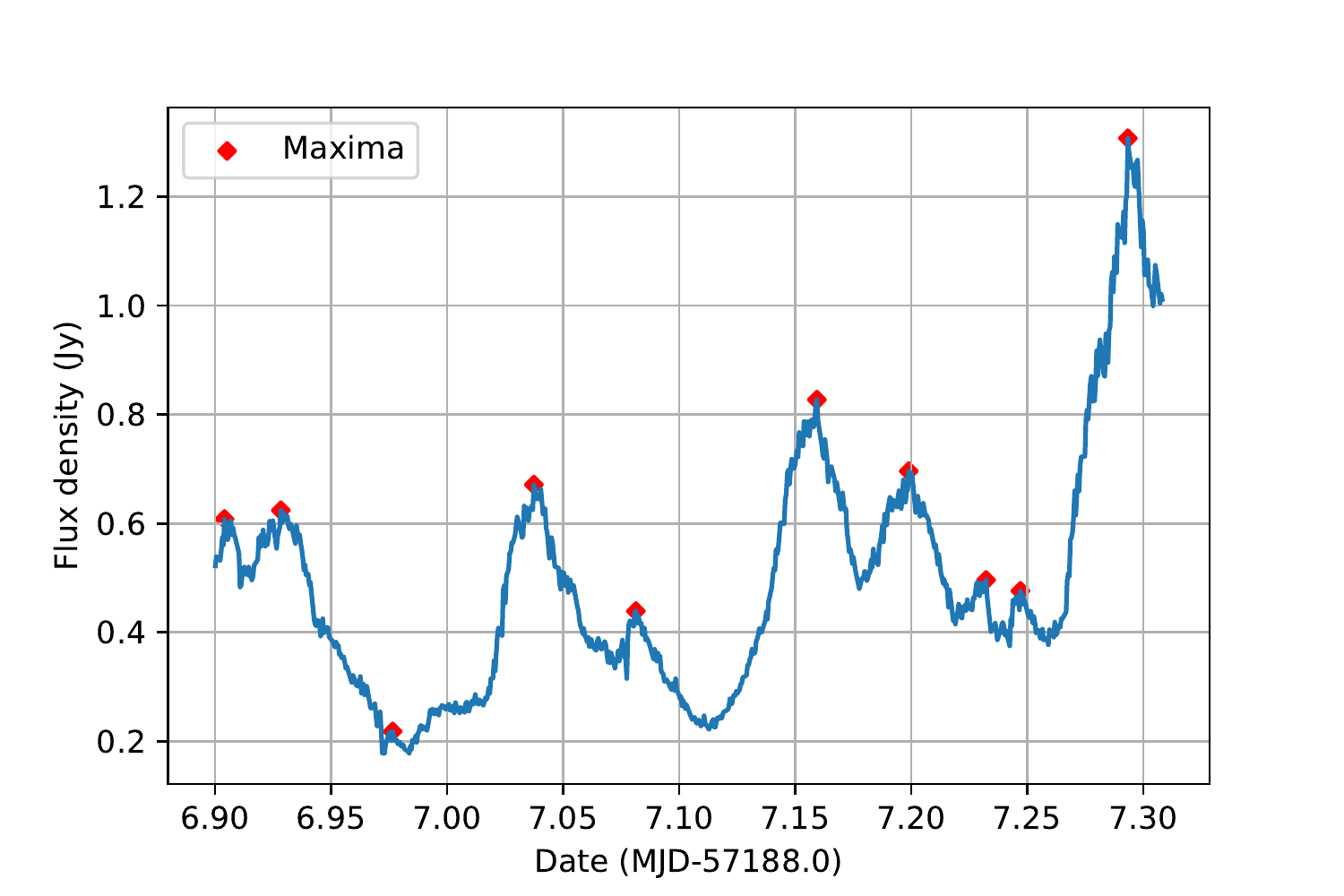, angle=0, width=7cm}\
\epsfig{file=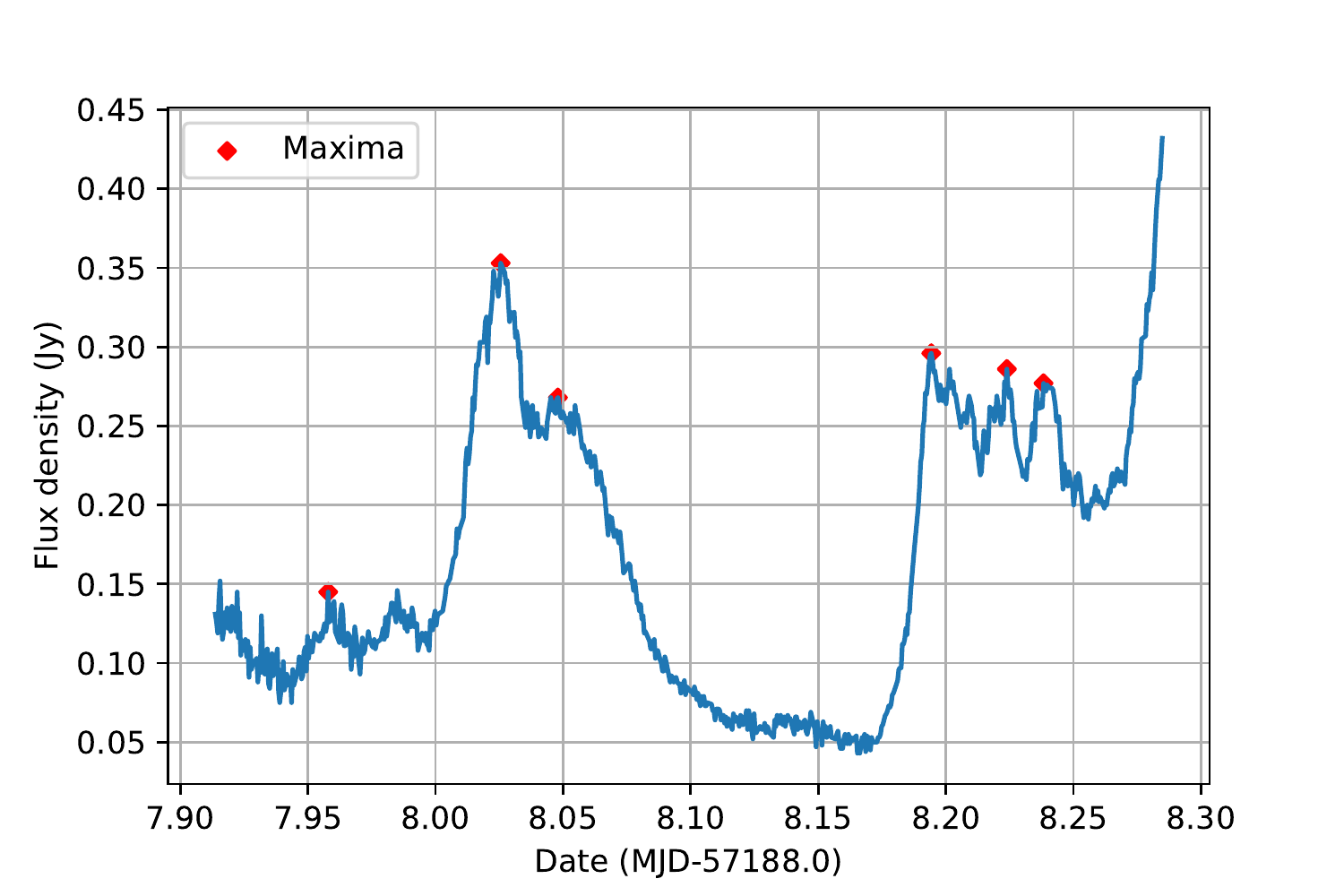, angle=0, width=7cm}\quad\epsfig{file=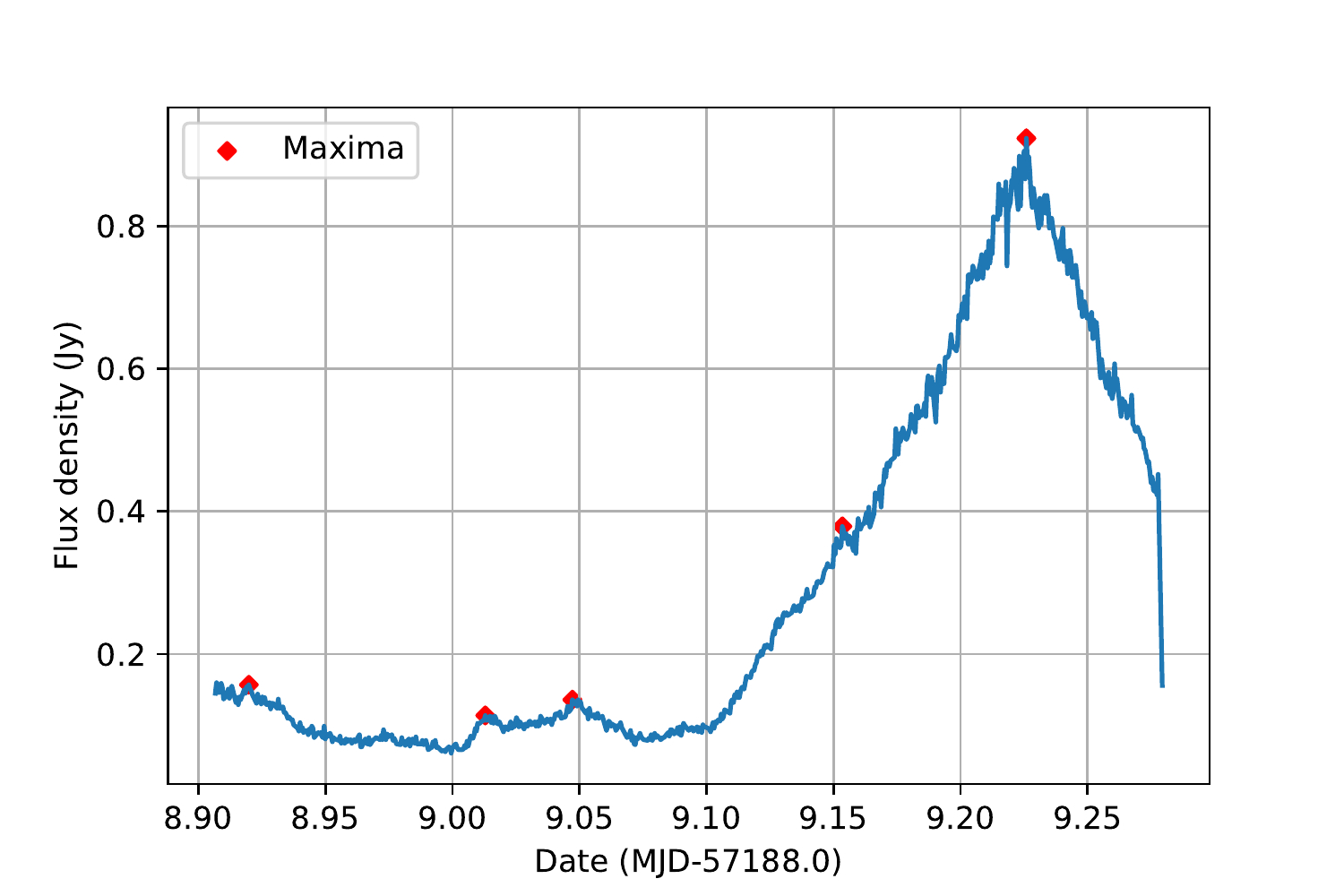, angle=0, width=7cm}\
\caption{Peaks in the AMI light curve automatically identified using the \textbf{scipy.signal.find\_peaks} package.}
\label{apppeaks}
\end{figure*}

\end{document}